\documentclass[useAMS,usenatbib]{mn2e}

\usepackage{graphicx}
\usepackage{epstopdf}

\def\lap{\;\rlap{\lower 2.5pt
\hbox{$\sim$}}\raise 1.5pt\hbox{$<$}\;}

\title[A Radiative Cooling Approximation]{An Efficient Radiative Cooling Approximation for Use in Hydrodynamic Simulations}
\author[J.\ C.\ Lombardi, W.\ G.\ McInally, and J.\ A.\ Faber]{James C.\ Lombardi Jr.$^{1}$\thanks{E-mail: jamie.lombardi@allegheny.edu}, William G.\ McInally$^{1}$, and Joshua A.\ Faber$^{2}$\\
$^{1}$Physics Department, Allegheny College, 520 North Main Street, Meadville, PA 16335, USA\\ 
$^{2}$School of Mathematical Sciences and Center for Computational Relativity and Gravitation, Rochester Institute of Technology,\\Rochester, NY 14623, USA}

\begin{document}

\pagerange{\pageref{firstpage}--\pageref{lastpage}} \pubyear{2014}

\maketitle

\label{firstpage}

\begin{abstract}
To make relevant predictions about observable emission, hydrodynamical simulation codes must employ schemes that account for radiative losses, but the large dimensionality of accurate radiative transfer schemes is often prohibitive.
Stamatellos and collaborators introduced a scheme for smoothed particle hydrodynamics (SPH) simulations based on the notion of polytropic pseudo-clouds that uses only local quantities to estimate cooling rates.  The computational approach is extremely efficient and works well in cases close to spherical symmetry, such as in star formation problems.  Unfortunately, the method, which takes the local gravitational potential as an input, can be inaccurate when applied to non-spherical configurations, limiting its usefulness when studying disks or stellar collisions, among other situations of interest.  Here, we introduce the ``pressure scale height method,'' which incorporates the fluid pressure scale height into the determination of column densities and cooling rates, and show that it produces more accurate results across a wide range of physical scenarios while retaining the computational efficiency of the original method.  The tested models include spherical polytropes as well as disks with specified density and temperature profiles.
We focus on applying our techniques within an SPH code, although our method can be implemented within any particle-based Lagrangian or grid-based Eulerian hydrodynamic scheme.
Our new method may be applied in a broad range of situations, including within the realm of stellar interactions, collisions, and mergers. 
\end{abstract}

\begin{keywords}
hydrodynamics -- radiative transfer -- protoplanetary discs -- circumstellar
matter -- stars: formation -- stars: peculiar.
\end{keywords}

\section{Introduction}

Modeling radiative cooling and transport is an extremely difficult task for numerical hydrodynamic simulations of astrophysical systems, whose dimensionality severely limits what can be accomplished in a reasonable amount of time.  Indeed, throughout a 3-dimensional volume, one must keep track of absorption, scattering, and emission processes at every moment in time, for photons whose intensity can vary with respect to both the direction of travel and the photon frequency.
This task lies outside the capabilities of most, if not all, numerical simulation codes, including both grid-based Eulerian ones and particle-based Lagrangian codes, though the problem is particularly severe for the latter.  As a result, virtually every major numerical code that undertakes a study of radiative transfer for dynamically evolving configurations makes use of simplifying assumptions.

The most sophisticated radiation transport techniques have generally been implemented in Eulerian codes.  These tend to include
flux-limited diffusion schemes \citep{1981ApJ...248..321L},
%
%
schemes that involve ray tracing the propagation of photons, or some combination of both.  Examples of codes capable of multigroup flux-limited diffusion include FLASH  \citep{2000ApJS..131..273F}, ZEUS \citep{2001ApJS..135...95T}, ORION \citep{2007ApJ...656..959K}, and RAMSES \citep{2011A&A...529A..35C}, with FLASH among others also allowing the option of a hybrid characteristic/diffusion method as well \citep{2006A&A...452..907R}.  In general, these techniques are well-developed, but far from trivial to implement and potentially a significant contributor to the computational resources required to perform a simulation.

Radiative transfer and cooling has been implemented at a much more basic level in Lagrangian simulations, to the extent it is modeled at all, and we focus our attention almost exclusively in that context.  One obvious problem is the irregular distribution of the matter components, which makes following the radiation flow throughout space substantially  more difficult.
Much of the work to date for Lagrangian problems has been specifically developed in terms of 
Smoothed particle hydrodynamics, hereafter SPH \citep{1977AJ.....82.1013L,1977MNRAS.181..375G}, the most popular particle-based hydrodynamics approach in  astrophysics, which is also
used widely throughout the fluid mechanics community \citep{2012AnRFM..44..323M}.  In  astrophysics, it has been used to investigate a broad range of dynamical processes, including stellar collisions (see, e.g. \citealt{2014ApJ...786...39N}) and star formation (e.g., \citealp{2014MNRAS.439.3039L}).  The method works by breaking down a fluid configuration into a large number of particles, each of finite, non-zero volume, which are allowed to interpenetrate and overlap.  Hydrodynamic and thermodynamic quantities are determined by computing local averages.  For reviews on the use of SPH  in astrophysics, see \citet{2009NewAR..53...78R,2010ARA&A..48..391S}.  

One approach for modeling radiative cooling, introduced by \citet[hereafter SWBG]{2007A&A...475...37S}, uses local quantities to help approximate the surrounding fluid configuration with polytropic pseudo-clouds.  By using the gravitational potential and density as estimators for the location of a particle relative to the surface, one may approximate the local column density and optical depth, and use these as inputs into the radiative cooling rate.  
The method does hinge on the assumption that the fluid configuration is roughly spherical, which can be a significant limitation.  In simulations of dynamical collisions and interactions between stars, for example, the use of gravitational potential as a tracer leads to incorrect results when a companion star is introduced, and during and after a stellar interaction, when the material may take on a disk-like or other non-spherical form.
Indeed, for non-spherical objects, the gravitational potential is non-constant at the surface, and the method may substantially err in locating regions with smaller optical depths.   

Even for spherical configurations, it is difficult for codes that use local quantities to approximate column densities to properly compute cooling rates in optically thick regions, since diffusion rates require more detailed information about thermal energy gradients.  To alleviate this problem, \citet{2009MNRAS.394..882F} introduced a hybrid method combining the pseudo-cloud methods found in SWBG with a particle-neighbor based flux-limited  diffusion term.  They found more accurate thermalization of matter in optically thick regions, while maintaining the performance of the cooling technique for spherical configurations.

Noting the issues with fluid geometries, attempts have been made to incorporate gravitational accelarations and potentials into column density models designed specifically for disks.  
\citet{2012MNRAS.426.1061Y}, for instance, have introduced a method that uses the distance to a central source and the vertical component of the gravitational acceleration to estimate column densities.
The method is computationally efficient and  accurate for disk simulations, but performs poorly for roughly spherical objects, including bound clumps within a disk or, presumably, stellar configurations, as noted by the authors themselves.

Here, we introduce the ``pressure scale height method,''  in which 
the role of the gravitational potential in the
method introduced in SWBG is instead played by
the local pressure scale height.  In retaining the local nature of the calculation, we retain the overall efficiency of the method, but in a form that may more naturally be applied 
 to dynamically evolving fluid configurations of arbitrary geometry.
In doing so, we end up capturing many of the best features of the original SWBG method {\em as well as} variants that perform better on disk-like configurations, all without additional computational complexity.
  This method has now been incorporated into the {\tt Starsmasher} SPH code developed by the authors \citep{2010MNRAS.402..105G}.

Our work is motivated by the long history of SPH for studies of merging stars.   One of us (J.C.L.) has used SPH to model the system V1309 Sco as a stellar merger taking place within a common envelope, which resulted in substantial amounts of dynamically ejected material and a strongly differentially rotating stellar remnant \citep{2014ApJ...786...39N}.   We are also currently using the method to study the system V838 Monocerotis, which has long been suggested as a potential ``mergeburst'' source, in which two stars undergo several dynamical interactions before eventually coalescing to form a single remnant \citep{2006A&A...451..223T}.  Our calculations demonstrate that SPH can follow stars through several interactions while following the evolution of the merger, the ejecta from the system, and the resulting light curves.
This work also has important applications for studies of accretion disks around kicked black holes, which has been studied using our code \citep{2012ApJ...745...71P} and others (see, e.g., \citealp{2010PhRvD..81d4004A,2010MNRAS.404..947C,2010MNRAS.401.2021R}).

To understand the numerical impetus behind our techniques, it is important to consider the factors that determine the overall efficiency of an SPH code.   Roughly speaking, there are three timescales for computational routines, and the longest of these will determine the code's overall performance:
loops over each SPH particle are extremely fast, while loops over each particle and its neighbors may be roughly a hundred times slower.  Pairwise interactions between every particle are the slowest, and generally require special purpose hardware, e.g., GPUs \citep{2007astro.ph..3100H,2010MNRAS.402..105G} or a GRAPE chip (see \citealp{2001NewA....6...79S}.


Full radiative transport would exceed even the longest of these timescales if performed exactly, and direct calculations of column densities potentially fall into this category as well. 
Local calculations, on the other hand, have the potential to scale like the number of particles and can be much more efficient, assuming we can approximate the conditions through which photons will travel, in particular the column densities   between a particle and the surface of the fluid.
For a dynamically perturbed configuration, the local density will still be a useful input into an approximation, 
but the pressure scale height, which is already calculated for each particle to determine hydrodynamic forces, serves as a much better tracer of a particle's depth within the configuration than gravitational potential, as introduced in the original  SWBG method.
 
 Our paper is organized as follows.
In Section~\ref{sec:method}, we work through our revised method, along with related simplifications that allow for greater computational efficiency when calculating radiative losses.  In Sec.~\ref{sec:tests}, we present several tests of our new method.  Finally, in Sec.~\ref{sec:discussion}, we discuss examples of when our new method will be useful in calculating the electromagnetic signatures of various events.

\section{Pseudo-cloud methods for radiative transfer in dynamical configurations}\label{sec:method}

For a review of the SPH implementation used within the {\tt Starsmasher} code, we refer the reader to previous and current works in which it has been described, particularly  \citet{2010MNRAS.402..105G}.  Here, we focus instead on the energy equation implementation, as this represents our primary modification to the codebase.  If we define $u_i$ to be the specific internal energy of the $i$th SPH particle, our energy  equation has traditionally taken the form  
\begin{equation}
\left.\frac{du_i}{dt}\right|_{\rm HYDRO} = \left.\frac{du_i}{dt}\right|_{\rm Press}+\left.\frac{du_i}{dt}\right|_{\rm AV}
\end{equation}
a sum of contributions involving pressure forces exchange between pairs of neighboring particles and contributions from the artificial viscosity treatment used to model shocks by damping velocities and heating material in converging flows.  Our AV scheme uses the popular ``Balsara switch'' technique \citep{1995JCoPh.121..357B}, which suppresses dissipation in rotating flows as opposed to converging ones.  Here, we wish to add a radiative term, following the approach found in SWBG, but with some variations designed to greatly increase the accuracy of the method for non-equilibrium and particularly non-spherical configurations.

\subsection{The pseudo-cloud}

In this subsection, we follow the derivation of \S2.2 of \citet{2007A&A...475...37S}, except that we use pressure scale height $H_{\rm P}\equiv P/ |{\vec \nabla} P|$ rather than gravitational potential $\psi$ as the indicator of the depth of an SPH particle $i$ within the system.
Given that the hydrodynamic acceleration $\vec a_{h}=-{\vec \nabla} P/\rho$, where $\rho$ is density, is already calculated for each particle within any SPH code, the scale height at particle $i$ can be conveniently determined from
\begin{equation}
H_{{\rm P},i}= \frac{P_i}{\rho_i |{\vec a}_{{\rm h},i}|},
\end{equation}
where $\rho_i$ and $P_i$ are, respectively, the local density and pressure at particle $i$.  
The hydrodynamic acceleration ${\vec a}_{{\rm h},i}$ of particle $i$ is calculated by the appropriate sum over neighboring particles: in the case of the {\tt Starsmasher} SPH code, this summation is given by equations (A11) and (A12) of \cite{2010MNRAS.402..105G}, while in the case of the {\tt GADGET-2} SPH code, the summation is given by equation (7) of \cite{2005MNRAS.364.1105S}.  We note that the hydrodynamic acceleration excludes any contribution due to gravity.  Furthermore, we do not include a contribution from the artificial viscosity, as the calculated scale height should be independent of the velocity field.

We use the same general approach and notation as in SWBG.  For
example, $R=\xi R_{\rm o}$ is the radius of an SPH particle within a pseudo-cloud, where $\xi$ is the dimensionless radius and $R_{\rm o}$ is the scale-length.  The structure of the pseudo-cloud is described by the usual Lane-Emden function $\theta(\xi)$ through the polytropic index $n$.  However, the pseudo-cloud need not be in hydrostatic equilibrium.  Instead, the central pseudo-cloud density $\rho_{\rm c}$, scale-length $R_{\rm o}$, and central density $T_{\rm c}$ are chosen such that the actual density $\rho_i$, pressure scale height $H_{{\rm P},i}$, and temperature $T_i$ at the location of particle $i$ are replicated.  In particular,
\begin{eqnarray}
\rho_i \label{density} & = & \rho_{\rm c} \theta^n(\xi), \\
 H_{{\rm P},i} & = &-\frac{R_{\rm o}}{(n+1)}\frac{\theta(\xi)}{\theta^\prime(\xi)}, \label{scaleheight}\\
T_i & = &T_{\rm c} \theta(\xi), \label{temperature}
\end{eqnarray}
where $\theta^\prime\equiv {\rm d}\theta/{\rm d}\xi$.  In deriving equation (\ref{scaleheight}), we use
$P=K \rho^{1+1/n}=K \rho_{\rm c}^{1+1/n}\theta^{n+1}$.
Solving for $\rho_{\rm c}$, $R_{\rm o}$, and $T_{\rm c}$ yields
\begin{eqnarray}
\rho_{\rm c} & = & \rho_i \theta^{-n}(\xi), \label{rhoc} \\
R_{\rm o} & = & - (n+1)H_{{\rm P},i}\frac{\theta^\prime(\xi)}{\theta(\xi)}, \label{Ro}\\
T_{\rm c} & = & T_i \theta^{-1}(\xi). \label{Tc}
\end{eqnarray}

Thus, the column density on a radial path from particle $i$ to the boundary $\xi_{\rm B} R_{\rm o}$ of the pseudo-cloud is
\begin{eqnarray}
%
\Sigma_i(\xi) & = & \int_{\xi^\prime=\xi}^{\xi^\prime=\xi_{\rm B}} \rho_{\rm c} \theta^n(\xi^\prime) R_{\rm o} d\xi^\prime \nonumber\\
& = &
-\frac{(n+1)\rho_i H_{{\rm P},i}\theta^\prime(\xi)}{\theta^{n+1}(\xi)} \int_{\xi^\prime=\xi}^{\xi^\prime=\xi_{\rm B}} \theta^n(\xi^\prime)d\xi^\prime. \label{Sigmai}
\end{eqnarray}
From this, the pseudo-mean column density $\bar \Sigma_i$ is calculated from the mass-weighted average of $\Sigma_i(\xi)$ over all allowable $\xi$:
\begin{eqnarray}
\bar \Sigma_i &= &
\frac{\int_{\xi=0}^{\xi=\xi_{\rm B}} \Sigma_i(\xi) \theta^n(\xi)\xi^2 d\xi}
{\int_{\xi=0}^{\xi=\xi_{\rm B}} \theta^n(\xi)\xi^2 d\xi} \nonumber\\
& = &  \zeta^{\prime} \rho_i H_{{\rm P},i}, \label{barSigmai}
\end{eqnarray}
where
\begin{equation}
\zeta^\prime =
\frac{n+1}{\xi_{\rm B}^2 \theta^\prime_{\rm B}}
\int_{\xi=0}^{\xi=\xi_{\rm B}}
\int_{\xi^\prime=\xi}^{\xi^\prime=\xi_{\rm B}} \theta^n(\xi^\prime)d\xi^\prime\,
\frac{\theta^\prime(\xi)}{\theta(\xi)} \xi^2 d \xi \label{zetanprime}
\end{equation}
and $\theta^\prime_B =\theta^\prime(\xi_B)$.
The prime in $\zeta^\prime$ is simply so this dimensionless quantity is not confused with the corresponding quantity $\zeta$ from SWBG.  Values of $\zeta$ and $\zeta^\prime$ are listed in Table \ref{table:zeta}: neither are terribly sensitive to $n$, with $\zeta$ being the less sensitive.  Equation (\ref{barSigmai}) is eminently reasonable: $H_{{\rm P},i}$ is roughly the distance from particle $i$ to the surface, so one would expect the product of $\rho_i$ and $H_{{\rm P},i}$ to estimate the column density, up to a multiplicative constant of order unity.

\begin{table}
\center{
\begin{tabular}{ccc}
$n$	&	$\zeta$	&	$\zeta^\prime$	\\
\hline\hline
0	&	0.381 & 0.830 \\
1	&	0.376 & 0.965 \\
1.5	&	0.372 & 1.014 \\
2	&	0.368 & 1.057 \\
3	&	0.350 & 1.136 \\
4	&	0.340 & 1.215
\end{tabular}
	\caption{Coefficient comparison.}
	\label{table:zeta}
}
\end{table}

Calculations of the optical depth $\tau_i$ and pseudo-mean optical depth $\bar \tau_i$ proceed analogously to that of the column density $\Sigma_i$ and pseudo-mean column density $\bar \Sigma_i$.  Given the Rosseland-mean opacity $\kappa_{\rm R}(\rho,T)$, we have
\begin{eqnarray}
\tau_i & = & \int_{\xi^\prime=\xi}^{\xi^\prime=\xi_{\rm B}} \kappa_{\rm R}(\rho_{\rm c}\theta^n(\xi^\prime),T_{\rm c}(\theta(\xi^\prime))\, \rho_{\rm c} \theta^n(\xi^\prime) R_{\rm o} d\xi^\prime \nonumber\\
& = & -(n+1)\rho_i H_{{\rm P},i}\frac{\theta^\prime(\xi)}{\theta(\xi)} \int_{\xi^\prime=\xi}^{\xi^\prime=\xi_{\rm B}}  \nonumber \\
&& \kappa_{\rm R}\left(\rho_i\left[\frac{\theta(\xi^\prime)}{\theta(\xi)}\right]^n,T_{\rm c}\left[\frac{\theta(\xi^\prime)}{\theta(\xi)}\right]\right) 
\left[\frac{\theta(\xi^\prime)}{\theta(\xi)}\right]^n d\xi^\prime \label{taui}
\end{eqnarray}
and
\begin{eqnarray}
\bar \tau_i & = & \frac{(n+1) \rho_i H_{{\rm P},i}}
{\xi_{\rm B}^2 \theta^\prime_{\rm B} }
\int_{\xi=0}^{\xi=\xi_{\rm B}} d\xi
\frac{\theta^\prime(\xi)}{\theta(\xi)} \xi^2 
\int_{\xi^\prime=\xi}^{\xi^\prime=\xi_{\rm B}} \nonumber \\
&& \kappa_{\rm R}\left(\rho_i\left[\frac{\theta(\xi^\prime)}{\theta(\xi)}\right]^n,T_i\left[\frac{\theta(\xi^\prime)}{\theta(\xi)}\right]\right)
\theta^n(\xi^\prime)d\xi^\prime\,. \label{bartaui}
\end{eqnarray}

As in SWBG, we find it convenient to define the pseudo-mean opacity as
\begin{equation}
\bar \kappa_i = \frac{\bar \tau_i}{\bar \Sigma_i},\label{barkappai}
\end{equation}
which, for an assumed $n$, is a function of only two variables, namely $\rho_i$ and $T_i$.
A densely populated table of pseudo-mean opacities can be generated by evaluating the double integral
\begin{eqnarray}
\bar \kappa_{\rm R}(\rho,T) & = & \frac{n+1}{ \zeta^\prime \xi_{\rm B}^2 \theta^\prime_{\rm B} }
\int_{\xi=0}^{\xi=\xi_{\rm B}} d\xi \frac{\theta^\prime(\xi)}{ \theta(\xi)} \xi^2  
\int_{\xi^\prime=\xi}^{\xi^\prime=\xi_{\rm B}} \nonumber \\
&&
\kappa_{\rm R}\left(\rho_i\left[\frac{\theta(\xi^\prime)}{\theta(\xi)}\right]^n,T_i\left[\frac{\theta(\xi^\prime)}{\theta(\xi)}\right]\right)
\theta^n(\xi^\prime)d\xi^\prime\,
\label{barkappaR}
\end{eqnarray}
for various density $\rho$ and temperature $T$ input values.
A value of $\bar \kappa_i$ for SPH particle $i$ is then obtained by using the particle density $\rho_i$ and temperature $T_i$ to interpolate among the data of the table.
Evaluating the pressure scale height $H_{{\rm P}.i}$ at the location of this particle then allows for the determination of the pseudo-mean column density $\bar \Sigma_i$ via equation (\ref{barSigmai}).  The pseudo-mean optical depth $\bar \tau_i=\bar \kappa_i \bar \Sigma_i$ is then used as an estimate of the number of mean free paths to infinity.  Because the pressure gradient typically points approximately along the direction in which optical depth increases most quickly,
the optical depth being calculated is approximately the minimum among all possible paths to infinity.

\subsection{Dynamical evolution}

Following SWBG, we can define a radiative cooling rate and background temperature for the simulation volume, where our equations are equivalent to theirs modulo the differences described above found by choosing pressure scale heights rather than the gravitational potential to determine location within the pseudo-cloud (see their Eqs.~24--25).
The radiative cooling rate is the quantity of ultimate interest:
\begin{equation}
\left.\frac{d u_i}{dt}\right|_{\rm RAD} = \frac{4 \sigma_{\rm SB}\left(T_0^4(\vec r_i)-T_i^4\right)}{\bar \Sigma_i^2 \bar \kappa_{\rm R}(\rho_i, T_i) + \kappa_{\rm P}^{-1}(\rho_i,T_i)}, \label{dudt}
\end{equation}
where $\sigma_{\rm SB}$ is the Stefan-Boltzmann constant, $\bar \kappa_{\rm R}$ is given by equation (\ref{barkappaR}) with $\zeta^\prime$ assumed to be 1.06, and $\kappa_{\rm P}$ is the Planck-mean opacity.  As in SWBG, we do not discriminate between the Rosseland-mean and Planck-mean opacities but instead use the same function for each.  We note that equation (\ref{dudt}) is an approximation designed to combine smoothly limiting expressions for the cooling rate in the optically thin (small $\bar \Sigma_i$) and optically thick (large $\bar \Sigma_i$) limits, and that the expression in the optically thick limit is itself an approximation (see WC and SWBG).

For a background temperature in our simulation, we follow their prescription and set
\begin{equation}
T_0^4 = T_{\rm min}^4 +\sum_* \frac{L_*}{16\pi \sigma_{\rm SB}|\vec{r}-\vec{r}_*|^2}
\end{equation}
where $T_{\rm min}$ provides a floor for the background temperature (SWBG choose $T_{\rm min}=10{\rm K}$), and the sum is over all nearby stars assumed to be irradiating the simulation volume.  In the case of the mergeburst model for V838 Monocerotis, for example, the B3 V companion with luminosity $L_*\approx 10^3 L_\odot$ and presumed distance $r_* \approx 300$ AU gives a value $T_0\approx 100$ K, with the choice of $T_{\rm min}$ being immaterial.  All arguments about the approximations above in the optically thin and thick cases presented in SWBG hold here as well, without modification.

For problems in which shock heating is the primary mechanism driving radiative losses, we adopt a slightly different approach than SWBG to handle the dynamics.  If we denote by $
\left.\frac{d u_i}{dt}\right|_{\rm HYDRO}$ the hydrodynamic evolution term for the internal energy combining pressure heating with contributions from an artificial viscosity scheme (see, e.g., Appendix A of \citet{2010MNRAS.402..105G} for our particular implementation or equation~(29) of SWBG), we typically do {\em not} include its effects when calculating the local thermalization timescale $t_{{\rm therm,}i}$.  Thus, unlike equations~(30) and (31) of SWBG, we define our equilibrium temperature as the {\em background temperature} $T_0$, and the thermalization timescale is determined solely by radiative losses:
\begin{eqnarray}
T_{{\rm eq}} &=&T_0\\
u_{{\rm eq},i}&=&u(\rho_i,T_{{\rm eq},i})\\
t_{{\rm therm,}i}&=&\frac{u_{{\rm eq},i} - u_i}{\left.du_i/dt\right|_{\rm RAD} }. \label{ttherm}
\end{eqnarray}
Finally, our update step includes a semi-implicit treatment of the radiative term only, not the hydrodynamic term:
\begin{eqnarray}
u_i(t+\Delta t) &=& u_i(t)\exp\left(\frac{-\Delta t}{t_{{\rm therm,}i}}\right) \nonumber \\
&&+u_{{\rm eq},i}\left[1-\exp\left(\frac{-\Delta t}{t_{{\rm therm,}i}}\right)\right] \nonumber \\
&&+ 
 \left.\frac{d u_i}{dt}\right|_{\rm HYDRO}\Delta t.\label{u_update}
\end{eqnarray}
There are admittedly trade-offs to our approach as compared to SWBG, which depend on the problem under consideration.  The advantages of our approach are twofold. First, it is simpler, since it does not require solving a non-linear equation for the equilibrium temperature $T_{{\rm eq}}$ as in equation~(30) of SWBG.  Moreover, defining the equilibrium temperature implicitly in terms of the radiative and hydrodynamic heating and cooling yields an expression that take multiple values in some circumstances.  For cases where $t_{{\rm therm,}i}\gg \Delta t$, this is not a problem, since the first-order approximation for the cooling rate is independent of $T_{{\rm eq}}$, but it can be important when $t_{{\rm therm,}i}\ll \Delta t$.  There are potential drawbacks to our approach as well.  Though our method and that of SWBG converge to the same solution when $t_{{\rm therm,}i}\gg \Delta t$, ours is less accurate in the opposite case when $t_{{\rm therm},i}$ is extremely small, since we essentially allow the hydrodynamic terms to overshoot the proper equilibrium temperature during a timestep.  In cases where $\left.\frac{d u_i}{dt}\right|_{\rm HYDRO}$ and $\left.\frac{d u_i}{dt}\right|_{\rm RAD}$ are sufficiently large and negative, it is theoretically possible for equation~(\ref{u_update}) to return a negative energy for a particle, though we find this does not occur in practice.  Given these limitations, we would recommend using the original timestepping approach of SWBG when one expects substantial emission from parts of the gas where the thermalization time will be small.  For dynamical simulations where the luminosity is dominated by higher-energy emissions from hot regions, and thermalization times across the vast majority of the configuration are long with respect to the timestep, the simplicity of our method makes it a fine choice.

\subsection{Summary of the pressure scale height method}

Before proceeding to tests, we wish to establish some basic principles regarding the pressure scale height method.  For those wishing to compare the methods side-by-side, Table~\ref{table:equations} lists an equation-by-equation comparison of the original method and our changes.  The basic implementation of our method is quite similar to that outlined in \S2.5 of SWBG.  In particular, in each iteration and for each SPH particle $i$, we
\begin{enumerate}
\item Use equation (\ref{barSigmai}) to determine the pseudo-mean column density $\bar \Sigma_i$ from the particle density $\rho_i$ and pressure scale height $H_{{\rm P},i}$.

\item Use pre-computed tables to interpolate a value for the 
the pseudo-mean opacity $\bar \kappa_{\rm R}(\rho_i, T_i)$, and determine the Planck-mean
opacity $\kappa_P(\rho_i, T_i)$.

\item
Calculate the radiative heating rate $du_i/dt|_{\rm RAD}$ from equation (\ref{dudt}) and the
 hydrodynamic heating rate
$du_i/dt|_{\rm HYDRO}$ from the usual SPH treatment.

\item Use equation (\ref{ttherm}) to determine the thermalization timescale $t_{{\rm therm},i}$.

\item Use equation (\ref{u_update}) to advance the specific internal energy $u_i$.
\end{enumerate}

We expect that both the original and our new methods should perform well when the underlying geometry of a fluid configuration is close to that of an isolated, spherical polytrope.  For binary stars that are near equilibrium, such as a system prior to a strong tidal interaction or just after a very weak one, our method should calculate cooling rates more accurately, since the original method in SWBG interprets the additional gravitational potential contribution from each star's companion as a systematic shift for particles to deeper depths within the star, whereas the modified method will remain invariant to leading order.  Given the simplified local treatment, both methods struggle to account for irradiation of each star by its companion, since this is not properly a local effect, and radiation is not actually transported across the system numerically.  Some aspects of this process will be captured by modifying the background temperature $T_0$, but it is clearly a case where more accurate and complicated treatments could be used.

For disk-like configurations, either primordial ones or those resulting from binary interactions, our method yields a much more physically motivated depth estimate for particles compared to the original one.  As we show in our tests below, the pressure gradient in a disk should in general provide a good estimate of depth within the disk and thus column density to the surface, to within factors of order unity, whereas the gravitational potential does not. This is true independently of the mass of the central gravitating object at the center of the disk, which spuriously changes depth estimates for the original method, but is basically ignored by our method, as it should be.

\begin{table}
\center{
\begin{tabular}{ccc}
Equation&	Corresponding&	 \\
in this	& equation	&	 \\
 paper	& in SWBG	&	Comment \\
\hline\hline
 (\ref{density}) &(10) & Same \\
(\ref{scaleheight}) &(11) &  $H_{{\rm P},i}$ replaces gravitational potential\\
(\ref{temperature}) &(15) &  Same\\
(\ref{rhoc}) &(13) &  Same\\
(\ref{Ro}) &(14) &  $R_{\rm o}$ now in terms of $H_{{\rm P},i}$\\
(\ref{Tc}) & (16) & Same\\
(\ref{Sigmai}) & (17) & $\Sigma_i$ now in terms of $H_{{\rm P},i}$\\
(\ref{barSigmai}) & (18) & $\bar \Sigma_i$ now in terms of $H_{{\rm P},i}$\\
(\ref{zetanprime}) & (19) & New dimensionless coefficient \\
(\ref{taui}) & (20) & $\tau_i$ now in terms of $H_{{\rm P},i}$\\
(\ref{bartaui}) & (21) &  $\bar \tau_i$ now in terms of $H_{{\rm P},i}$\\
(\ref{barkappai}) &(22) &  Same\\
(\ref{barkappaR}) &(23) &  New dependency on $\theta$\\
\end{tabular}
	\caption{Comparison of equations from SWBG and the present work.}
	\label{table:equations}
}
\end{table}

\section{Tests}\label{sec:tests}

\begin{figure}
	\centerline{\includegraphics[width=\columnwidth]{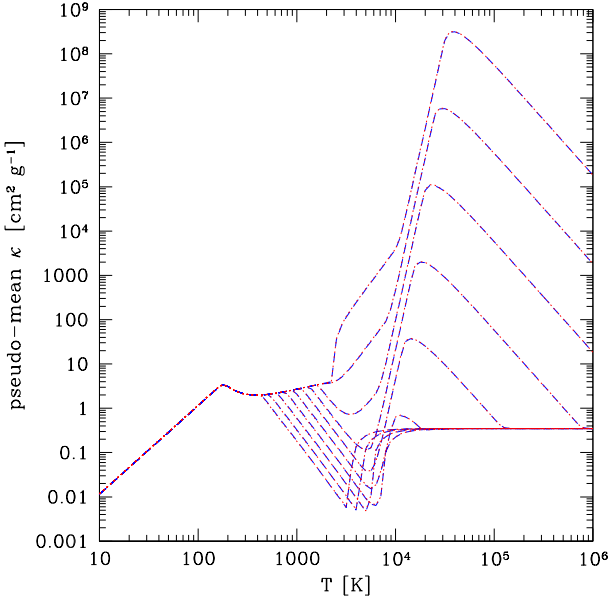}}
	\caption{Comparison of pseudo-mean opacity $\bar \kappa_{\rm R}(\rho,T)$ as calculated by SWBG (red dotted curves) and by equation (\protect\ref{barkappaR}) of the present work (blue dashed curves) with $n=2$.  Values typically agree to within a few percent.   
A curve is plotted for every two orders of magnitude in density from $10^{-18}$g cm$^{-3}$ (bottom) to 1 g cm$^{-3}$ (top).}  \label{pseudo}
\end{figure}

To test our method, we reproduce several diagnostics from \citet{2012MNRAS.419.3368W}, hereafter denoted WC, for spherical configurations and disks. First, however, we demonstrate that the comparisons to SWBG are fair and well defined.  In Figure \ref{pseudo}, we show a comparison of the pseudo-mean opacity $\bar \kappa_{\rm R}(\rho,T)$ as a function of temperature for fixed fluid density, calculated here from Equation (\ref{barkappaR}), and the corresponding result when using Equation (23) of SWBG (see their Figure 6 for the original presentation).  In both cases,  opacities taken from \citet{1994ApJ...427..987B} have been used for the comparison, although neither the SWBG method nor the present method is restricted to this opacity choice. 

The $\bar \kappa_{\rm R}$ values calculated from the two different approaches agree to within a few percent over at least 18 orders of magnitude in density.  
Given that the integrand in our equation (\ref{barkappaR}) for $\kappa_{\rm R}$ involves  $w^\prime\equiv(n+1)\theta^\prime/\theta$ while the integrand in the corresponding equation (23) of SWBG involves $w\equiv-(\theta^n/\phi)^{1/2}$ instead, 
it is at first surprising that the pseudo-mean opacities of the two methods agree so well.
Indeed, the $\xi$ dependence of the weighting function $w^\prime/\zeta^\prime$ from our integral is quite different from that of the weighting function $w/\zeta$ in the corresponding integral of SWBG, even though both of these functions represent the same dimensionless quantity, namely $-R_{\rm o}\rho/\bar \Sigma$.
%
%
The largest differences in these weighting functions occur near the center and the surface of a pseudo-cloud; however, these differences are suppressed in the outer integral by the multiplicative factor $\xi^2$, which approaches zero at the center, and by the inner integral, which approaches zero at the surface.

To understand the source of agreement in pseudo-mean opacities, we need to look back at the definitions of $\zeta^\prime$ and $\zeta$.
%
%
From equation 
 (\ref{zetanprime}), we see that $\zeta^\prime$ is proportional to a weighted average over all possible pseudo-clouds of an average of $w^\prime$ out to the surface of a pseudo-cloud.  That is, the integral for $\zeta^\prime$ looks just like that for $\bar \kappa_{\rm R}$, except that the later also includes a $\kappa_{\rm R}$ in the inner integrand.
Likewise, from the corresponding definitions in SWBG, we see that
$\zeta$ and $\bar \kappa_{\rm R}$ involve the same kind of integrals with both now involving $w$ instead of $w^\prime$.  So, in the end, although the weighting functions $w^\prime/\zeta^\prime$ and $w/\zeta$ being used to calculate $\bar \kappa_{\rm R}$ are different, the $\zeta^\prime$ and $\zeta$ normalize these functions in a way that gives a very similar result for the resulting pseudo-mean opacities in each approach.

\subsection{Column density estimates}

\begin{figure*}
	\includegraphics[width=0.84\columnwidth]{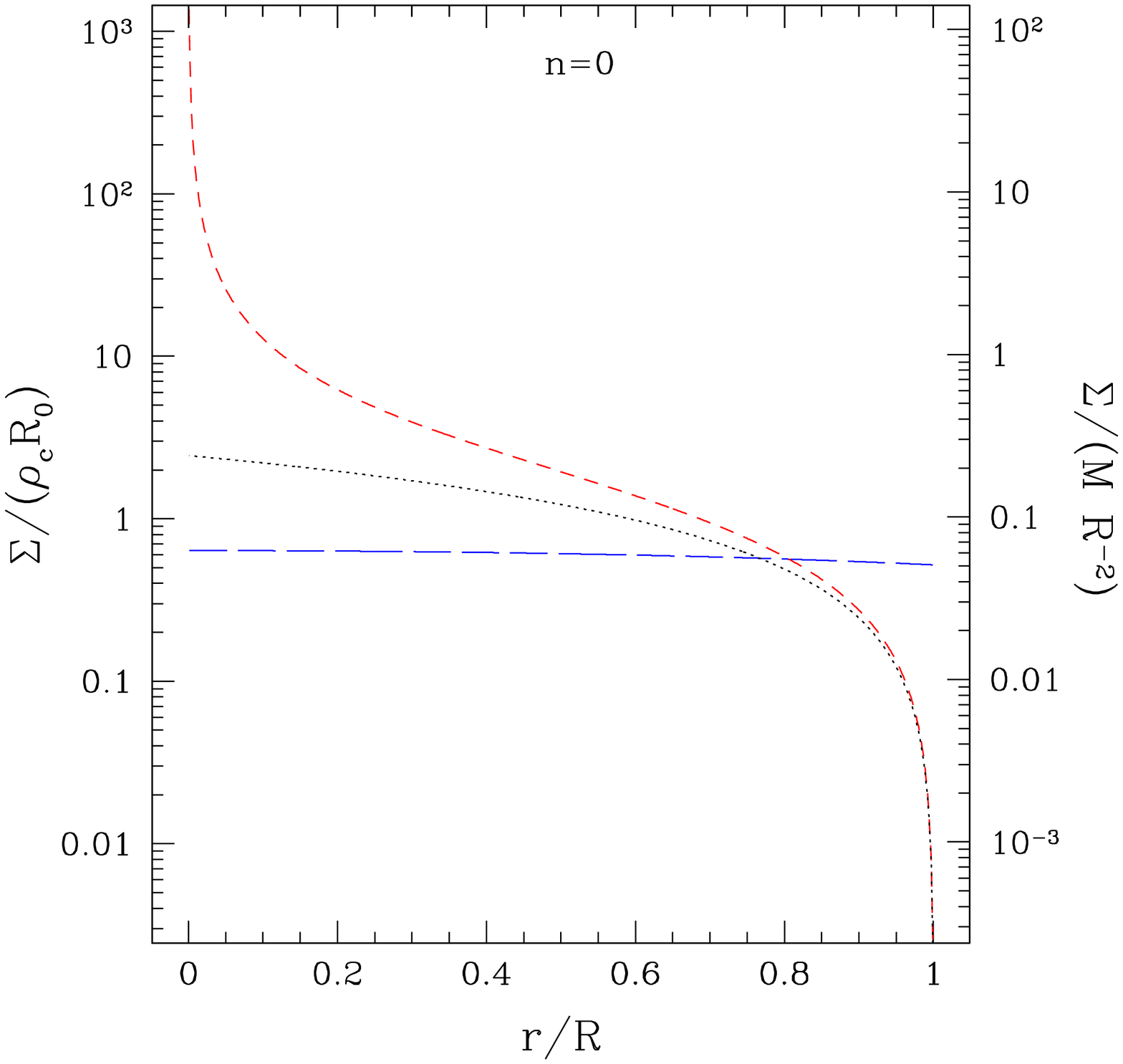}~~~~~
	\includegraphics[width=0.84\columnwidth]{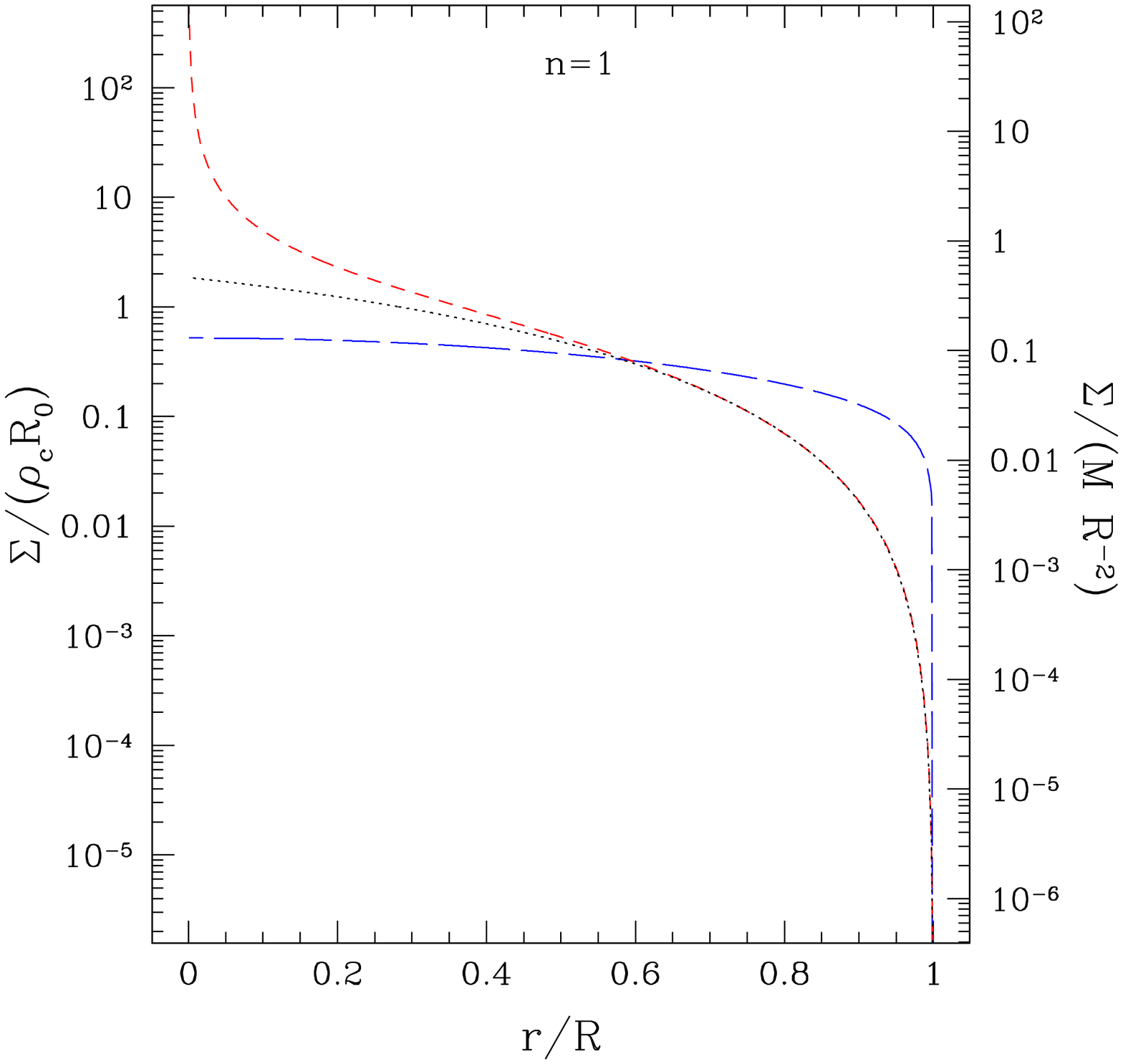}
	\includegraphics[width=0.84\columnwidth]{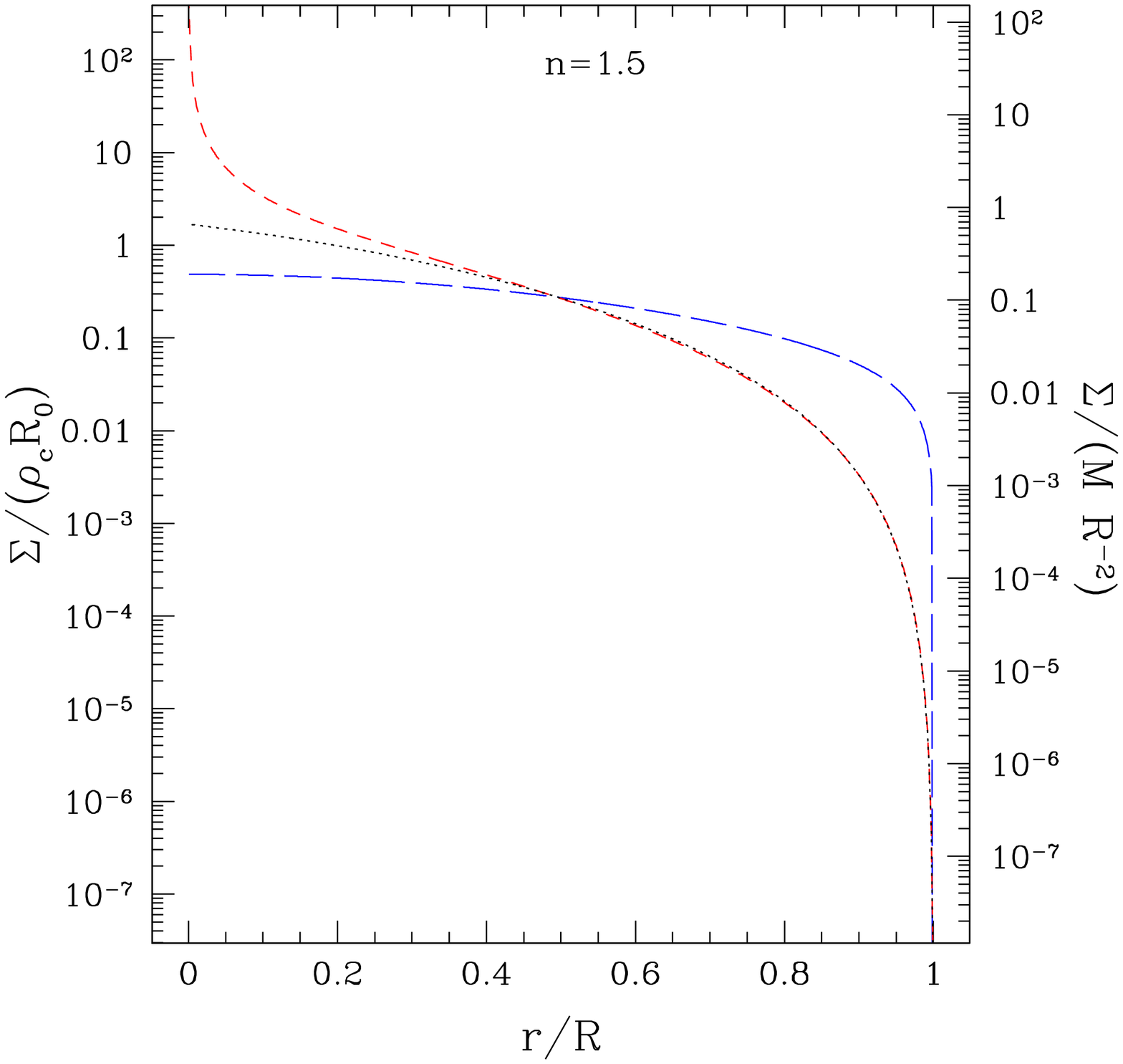}~~~~~
	\includegraphics[width=0.84\columnwidth]{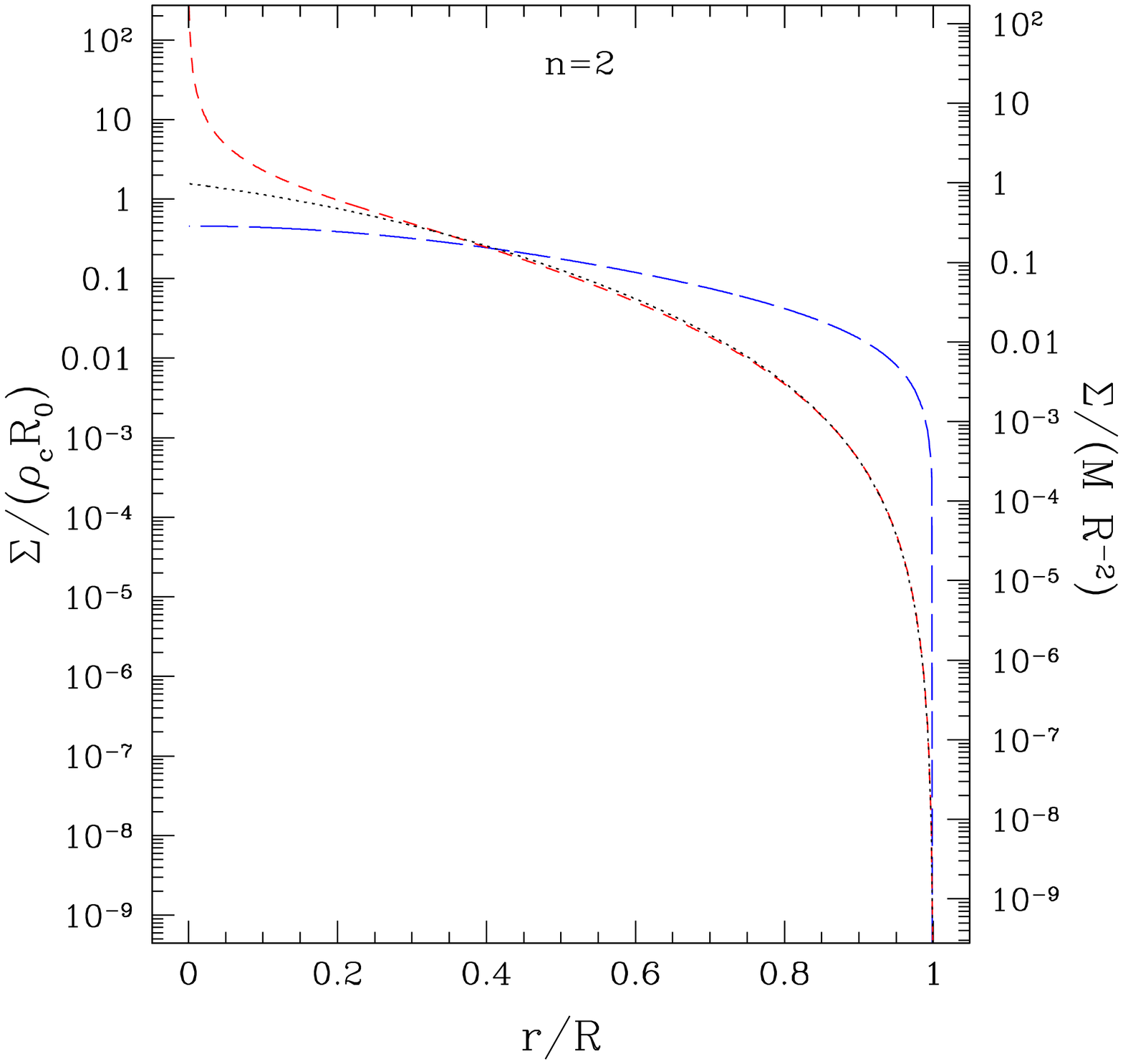}
	\includegraphics[width=0.84\columnwidth]{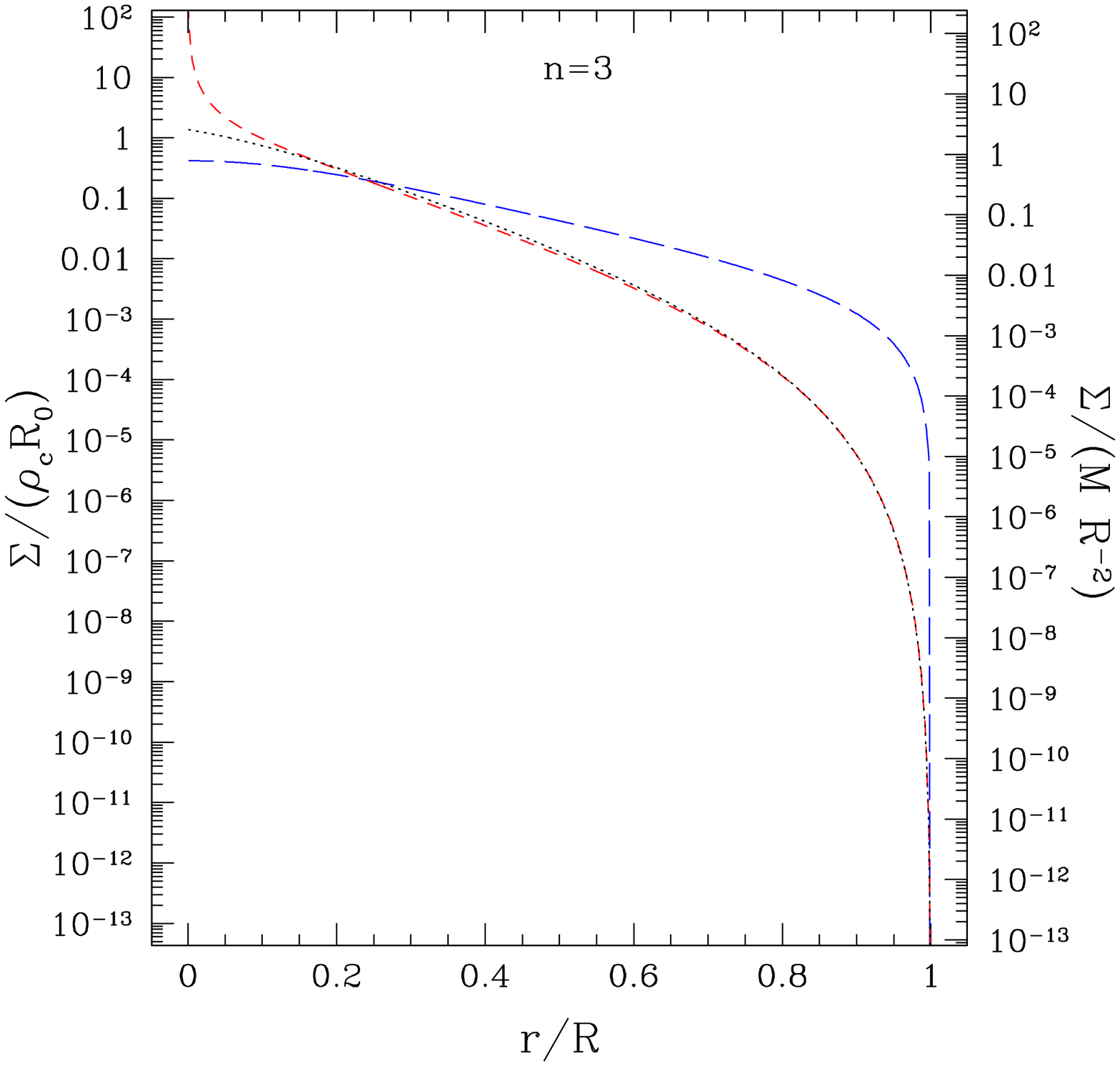}~~~~~
	\includegraphics[width=0.84\columnwidth]{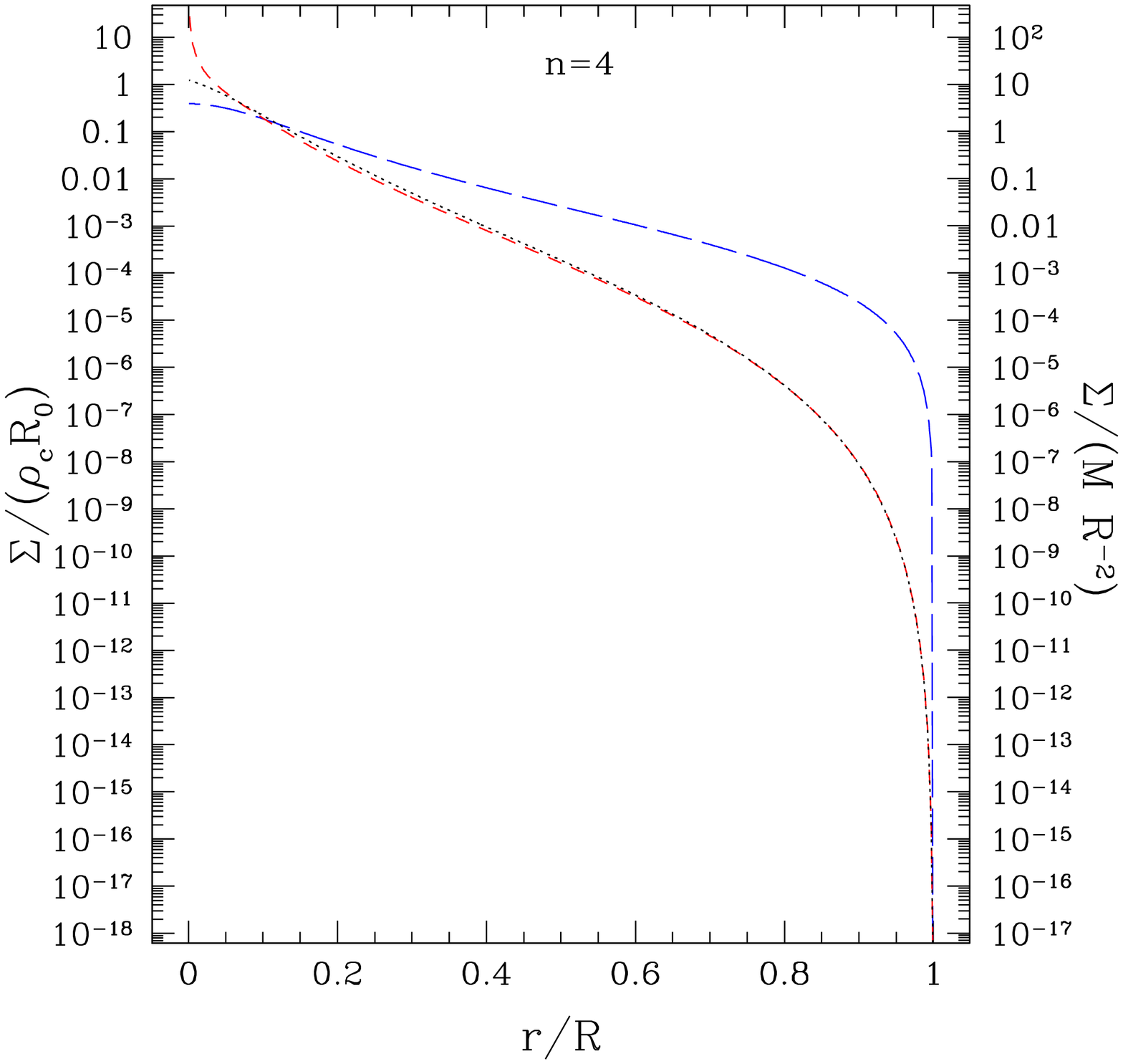}
	\caption{Column density $\Sigma$ or $\bar \Sigma$ (normalized to $\rho_{\rm c} R_0$ on the left axis and $MR^{-2}$ on the right axis) versus radius $r$ (normalized to the stellar radius $R$) for polytropes of index $n=0$, $1$, $1.5$, $2$, $3$, and $4$, as labelled near the top of each panel.  The dotted curve represents the actual $\Sigma=\int_\xi^{\xi_{\rm B}} \rho_{\rm c} \theta^n(\xi^\prime) R_{\rm o} d\xi^\prime$ obtained simply by integrating the density profile from $r$ to the surface at $R$.  The long dashed blue curve in each panel represents the pseudo-mean column density $\bar \Sigma$ as calculated by equation (18) of SWBG with $\zeta=0.368$, while the short dashed red curve represents that of our equation (\ref{barSigmai}) with $\zeta^\prime=1.06$.  To convert from the units of $\Sigma$ on the left axis to those on the right axis, we use that the stellar radius $R=\xi_B R_0$ and the stellar mass $M=4 \pi R_0^3 \rho_c \xi_B^2 |\theta^\prime(\xi_B)|$.  Note the good agreement of results, especially near the surface, between the actual column density and the approximation of our equation (\ref{barSigmai}).
} \label{Sigma}
\end{figure*}

Turning our attention to spherical stellar configurations,
Figure \ref{Sigma} compares the actual column density throughout a spherical polytrope to the pseudo-mean column densities calculated by equation (\ref{barSigmai}) and by the techniques of SWBG.
The data shown here are  consistent with those shown in Figure 2 of WC.  The dotted curve represents the column density $\Sigma$  calculated simply by integrating the density profile $\rho_{\rm c} \theta^n$ from the radius $r=\xi R_{\rm o}$ to the surface $R=\xi_{\rm B}R_{\rm o}$, namely $\Sigma=\int_\xi^{\xi_{\rm B}} \rho_{\rm c} \theta^n(\xi^\prime) R_{\rm o} d\xi^\prime$.  The red short dashed curve uses our equations (\ref{density}) and (\ref{scaleheight}) in equation (\ref{barSigmai}), with $\zeta^\prime =1.06$ regardless of $n$.  The computation using the SWBG approach employs their analogous equations (which can be identified easily with the help of Table \ref{table:equations}) with $\zeta^\prime =0.368$ regardless of $n$.  The $n$ dependence of $\zeta^\prime$ and $\zeta$ is purposely neglected because in an actual SPH simulation one
would not have readily available an index $n$ describing the overall structure of the system, even though the {\it local} quantities like density, pressure scale height, and gravitational potential could be easily determined.  In the hydrodynamic simulations we assume a fixed $\zeta^\prime$ (or $\zeta$), and so we do so here as well.  Similar results are shown for three of these cases in Figure 2 of WC, though we note the axes use slightly different scalings there.

Near the surface, where we would expect to find the photosphere in most problems of interest, our equation (\ref{barSigmai}) gives an accurate estimate of the actual column density.  As the center of the polytrope is approached, the pressure gradient approaches zero and hence both the pressure scale height and the pseudo-mean column density given by the present work diverge to infinity.  The optical depth of such regions would necessarily be overestimated.  Fortunately, provided only that the central region is well within the photosphere, the optical depth there will already be large, and, given the exponential falloff of radiation intensity with optical depth, the associated radiation leakage very small.  Therefore, the overestimate of column density will have little effect when confined to regions where $\tau>>1$.  It is much more important that the method can give accurate column density estimates in the vicinity of the photosphere.\footnote{More accurate column density estimates are possible if not only first derivatives but also second derivatives of pressure are used to evaluate the distance to the surface.  The extra complexity and computational workload are not necessary for the astrophysical problems, such as those involving optically thick outflows, that the present method is intended to treat.}

To test our approach for an astrophysically relevant case far from spherical symmetry, we consider the analytic disk models devised by WC.  These models assume a polytropic equation of state $P=K\rho^2$, where $K$ is a constant.
The density profile of the marginally self-gravitating model satisfies
\begin{equation}
\rho(R,z) = \rho_{\rm in}\left(\frac{R_{\rm in}}{R}\right)^3\left(2 \cos\frac{z}{H_0}-1\right),
\label{RhoDiskEquation}
\end{equation}
where $R$ is the cylindrical radius and $z$ is the height above the midplane.  The disk extends over $R_{\rm in}<R<R_{\rm out}$ and $|z|<H$, where $H=H_0 \cos^{-1}\frac{1}{2}=\pi H_0/3$ and $H_0=K^{1/2} (2\pi G)^{-1/2}$.  The density $\rho_{\rm in}=\rho(R_{\rm in},0)$ is the density on the midplane at the inner disk edge. The angular rotation speed satisfies $\Omega=(4\pi G\rho_{\rm in})^{1/2}(R_{\rm in}/R)^{3/2}$, corresponding to a Keplerian disk surrounding a central star of mass $4\pi \rho_{\rm in}R_{\rm in}^3$.

To obtain the column density, we simply integrate the density profile in the $z$ direction from $z$ to the surface at $H$:
\begin{equation}
\Sigma(R,z) = \rho_{\rm in}\left(\frac{R_{\rm in}}{R}\right)^3 H_0\left(\sqrt{3} - \frac{\pi}{3} +  \frac{z}{H_0} - 
   2 \sin\frac{z}{H_0}\right).
\label{SigmaDiskEquation}
\end{equation}
Strictly speaking, this is only an approximation to the most relevant column density for radiative transport, as the linear route that minimizes column density creeps to slightly larger values of $R$ due to the $1/R^3$ dependence of the density.  However, the error made by integrating in the direction exactly perpendicular to the midplane at, for example, $R=5R_{\rm in}$ is only $\sim$0.3\% at $z=0$ and even smaller at other $z$. 

\begin{figure}
	\centerline{\includegraphics[width=\columnwidth]{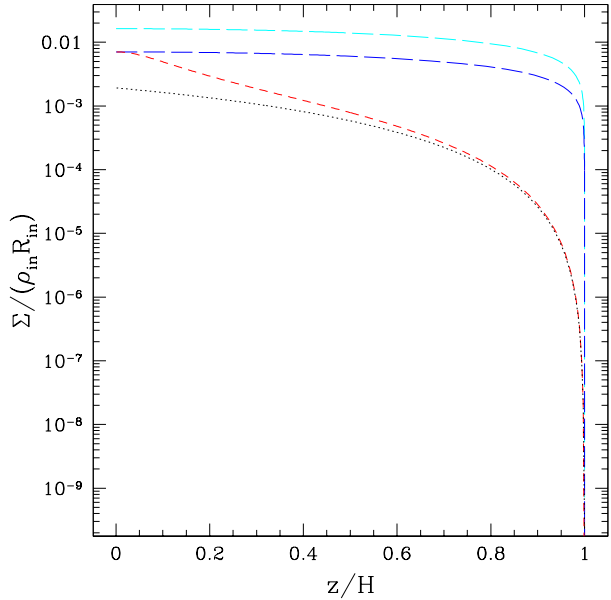}}
	\caption{Column density $\Sigma$ (normalized to $\rho_{\rm in} R_{\rm in}$) versus height $z$ (normalized to the disk height $H$) at $R=5R_{\rm in}$ in the disk of WC.  The dotted curve represents the actual column density $\Sigma$, as given by equation (\ref{SigmaDiskEquation}).  The short dashed red curve represents that of our equation (\ref{barSigmai}) with $\zeta^\prime=1.06$. The long dashed curves represent the pseudo-mean column density $\bar \Sigma$ as calculated by equation (18) of SWBG with $\zeta=0.368$: the top (cyan) curve uses the potential of the disk and star, while the next (blue) curve uses the potential of the disk only.  Note the good agreement of results, especially near the surface, between the actual column density and the approximation of our equation (\ref{barSigmai}).
} \label{SigmaDisk}
\end{figure}

We follow WC and adopt $R_{\rm out}=10 R_{\rm in}$ and $H_0=0.35 R_{\rm in}$.
Figure \ref{SigmaDisk} then compares the actual column density at $R=5R_{\rm in}$ to the pseudo-mean column densities calculated by equation (\ref{barSigmai}) and by the techniques of SWBG (see Figure 4 of WC).  As found by WC, the SWBG method overestimates the column density regardless of whether or not the contribution from the star is included in the potential: see the long dashed cyan and blue curves.\footnote{Evaluating the potential due to the disk involves complete elliptic integrals of the first kind: see e.g.\  \citet{1983CeMec..30..225L}.}
The red short dashed curve uses $\zeta^\prime =1.06$ in our equation (\ref{barSigmai}), which involves the pressure scale height, to obtain the pseudo-mean column density.  
Although the column density near the midplane is overestimated, we note that---in contrast to the centers of the polytropes considered in Figure \ref{Sigma}---the column density $\bar \Sigma$ remains finite even as $z\rightarrow0$ because the pressure gradient does not vanish due to the $R$ dependence of the density $\rho$.   
Furthermore, our results nicely match the column density near the edge of the disk, which is where we would expect to find the photosphere\footnote{We use the term ``photosphere" to mean the surface of optical depth $\tau=1$ even when that surface is non-spherical.} in many problems of interest.  

\subsection{Cooling rate estimates}

\begin{figure}
	\includegraphics[width=0.84\columnwidth]{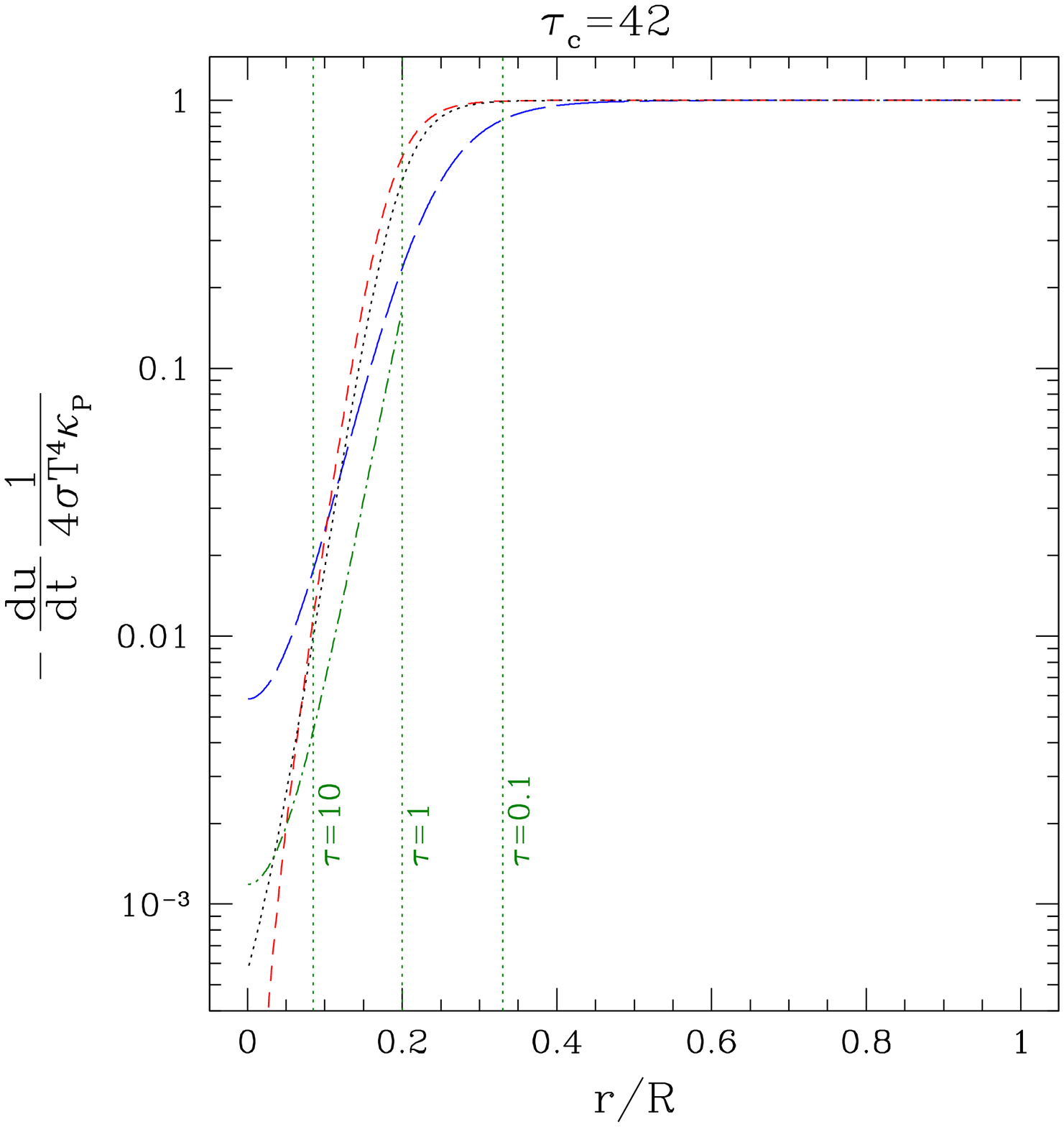}
	\includegraphics[width=0.84\columnwidth]{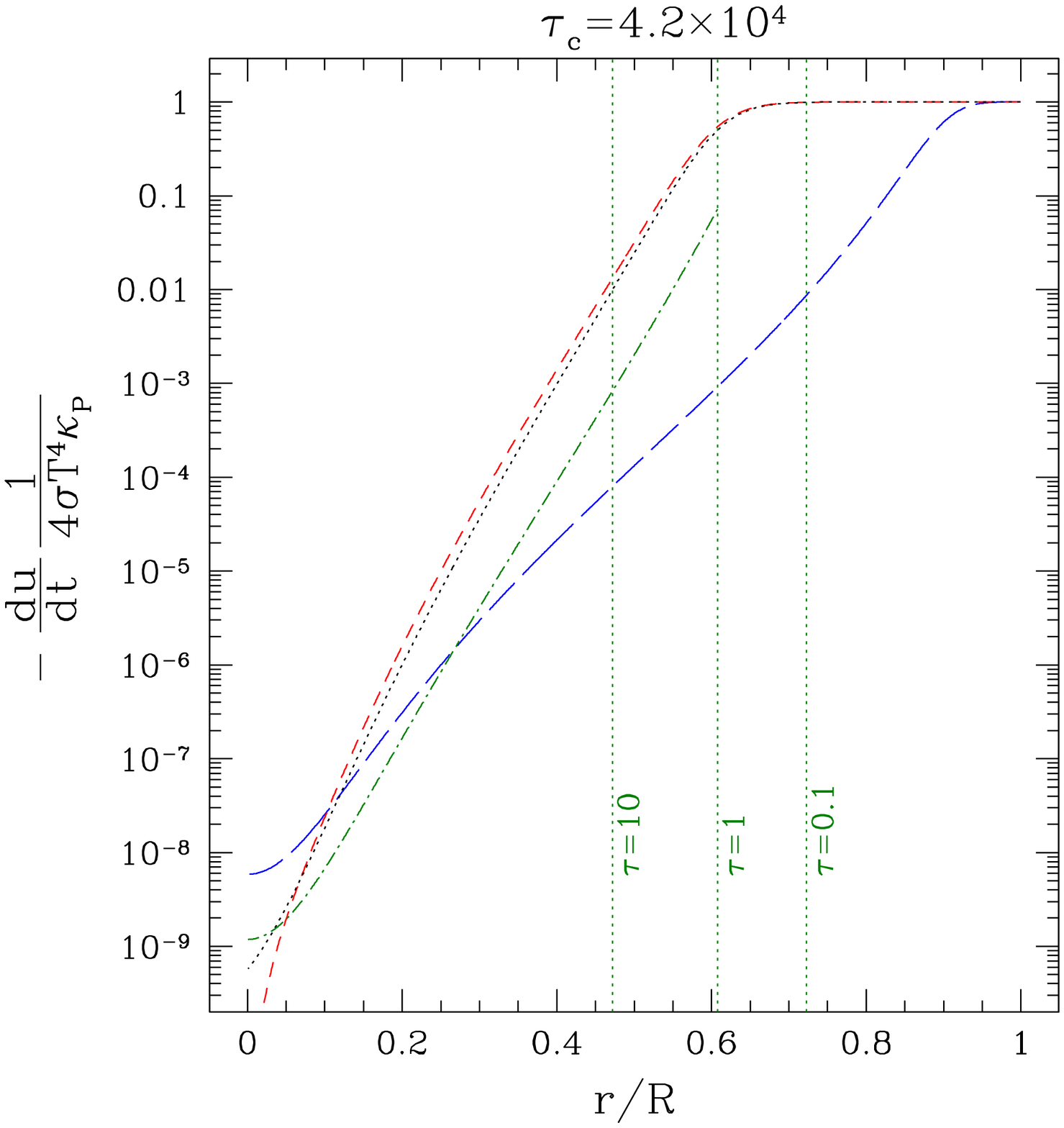}
	\includegraphics[width=0.84\columnwidth]{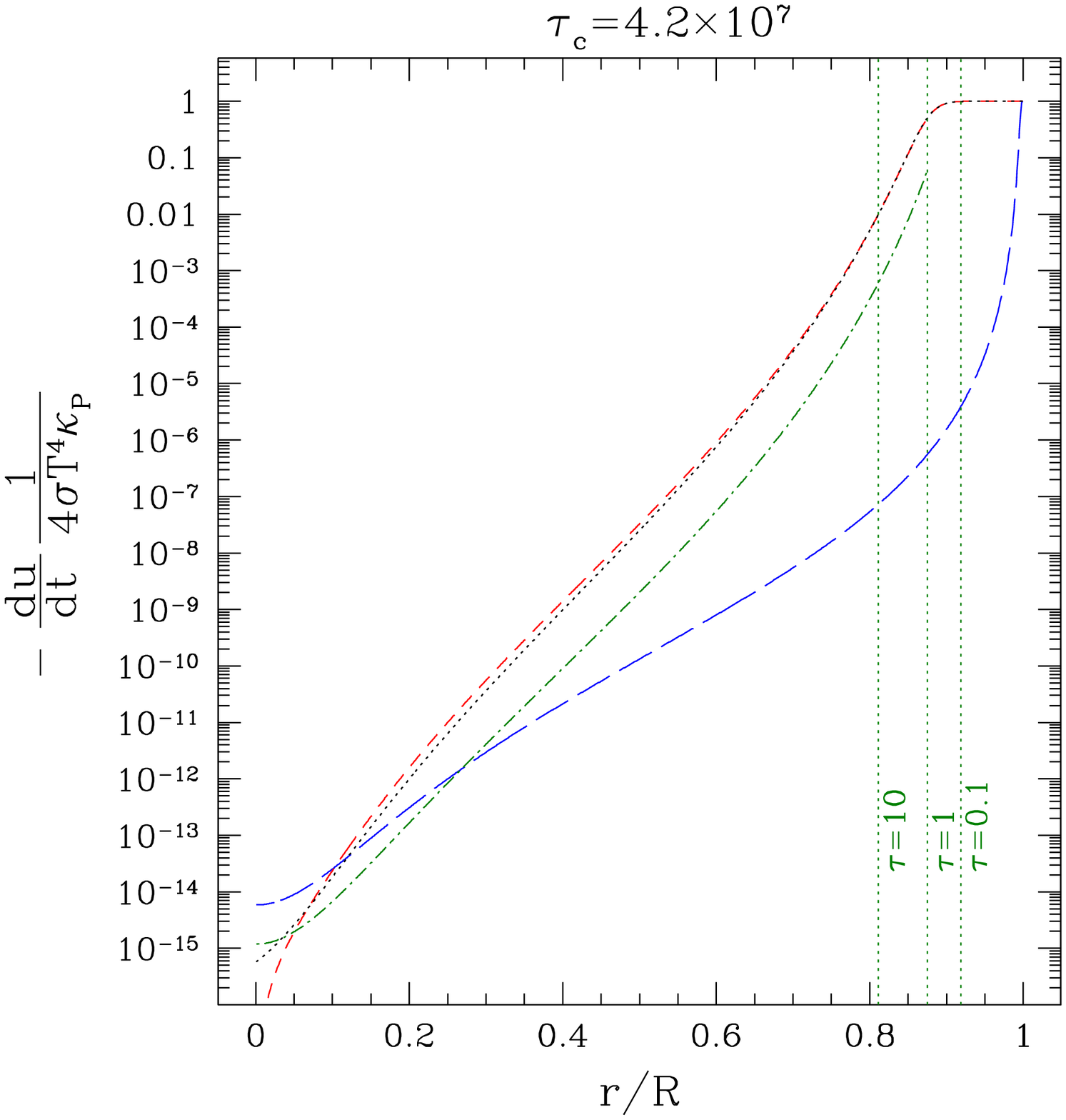}
	\caption{The radiative cooling rate $-du/dt$ (normalized to the optically thin result $4\sigma T^4 \kappa_{\rm P}$) versus radius $r$ (normalized to the stellar radius $R$) in an $n=4$ polytrope of constant opacity and with a central optical depth $\tau_{\rm c}$ of (a) 42, (b) $4.2\times 10^4$, and (c) $4.2\times 10^7$.  Line types are as in Figure \ref{Sigma}, with an additional green dot-dashed curve showing the diffuion limit result, assuming a temperature profile $T\propto P/\rho$, at true optical depths $\tau>1$. Vertical dotted lines show the locations where $\tau=$ 0.01, 1, and 10.
%
%
} \label{fig:dudt}
\end{figure}

Having established that our method yields excellent agreement for computed versus physical column densities for both polytropes and disk models, especially in regions near the surface of the fluid, we turn our attention to the predicted cooling rates, since this is the main use of the method.
Figure \ref{fig:dudt} compares the cooling rate $-du/dt$ within a constant opacity $n=4$ polytrope as determined by equation (\ref{dudt}) in three different ways.  The dotted curve is the result for which the actual column density is used.  The red short dashed curve uses equation (\ref{barSigmai}) to estimate the column density, while the blue long dashed curve uses the estimate of SWBG.  For consistency with Figure 6 of WC, we choose $T_0=0$ and $\kappa_{\rm P}=\kappa_{\rm R}=\tau_{\rm c}/\Sigma_{\rm c}=12.0\tau_{\rm c}\rho_{\rm c}^{-1}R^{-1}$, where $\tau_{\rm c}$ is the desired central optical depth and $\Sigma_{\rm c}$ is the actual column density at the center.  There is no need to assume a temperature profile for our curves except for the diffusion limit result, for which we use $T\propto P/\rho$ to calculate the radiative flux ${\bf F}=-4 \nabla(\sigma T^4)/(3\kappa\rho)$ and ultimately the cooling rate $du/dt=-\rho^{-1} \nabla\cdot {\bf F}$ in the optically thick limit.

As compared to Figure 6 of WC, our Figure \ref{fig:dudt} covers a larger range of radii, and we plot the cooling rate $-du/dt$ normalized to the optically thin limiting expression $4\sigma T^4 \kappa_{\rm P}$.  Frame (a) represents a generalization of the scenario presented in Figure 6(a) of WC.\footnote{We note that the right hand side of WC's equation (7) is missing a multiplicative factor of 4.  As a result, the corresponding curve in the figures of WC are closer to the diffusion limit result than what we find in Figure \ref{fig:dudt}.}  Frames (b) and (c) are more representative of cases where there is an optically thick outflow at densities that are orders of magnitude less than the maximum density in the system.
%
%
As compared to the original method of SWBG, our method is better able to reproduce the actual $du/dt$, with the overall agreement improving as the central optical depth increases and the photosphere moves closer to the surface.  Further improvements for actual SPH realizations of these polytropic stars, and the disks discussed below, could potentially be attained by adopting a hybrid pseudo-cloud/flux-limited diffusion scheme like that of \citet{2009MNRAS.394..882F}. However, such enhancements would have limited effect on the observable emission from systems with long thermal timescales, especially since the affected regions are concentrated well within the photosphere.

\begin{figure}
	\includegraphics[width=0.84\columnwidth]{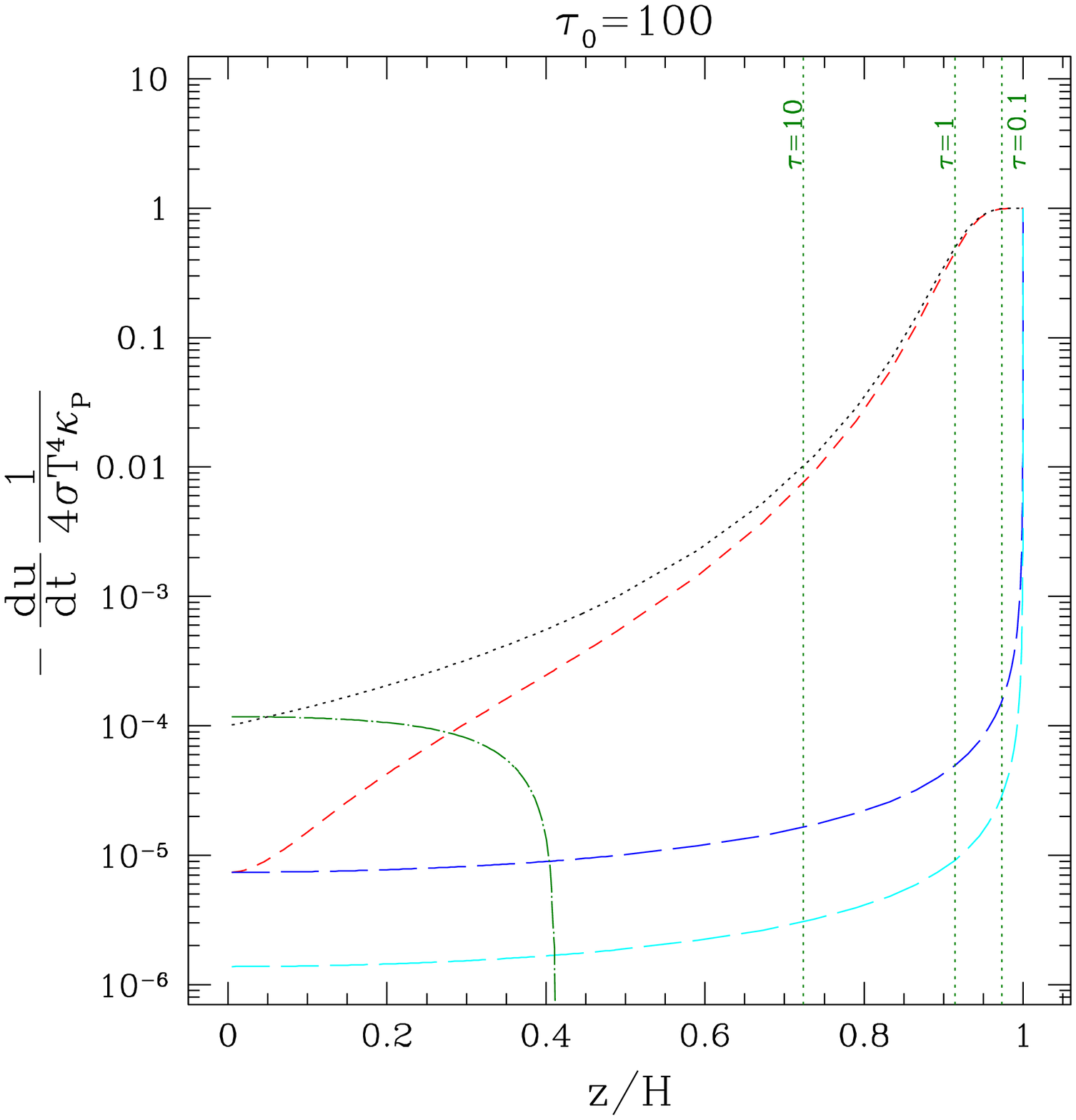}
	\includegraphics[width=0.84\columnwidth]{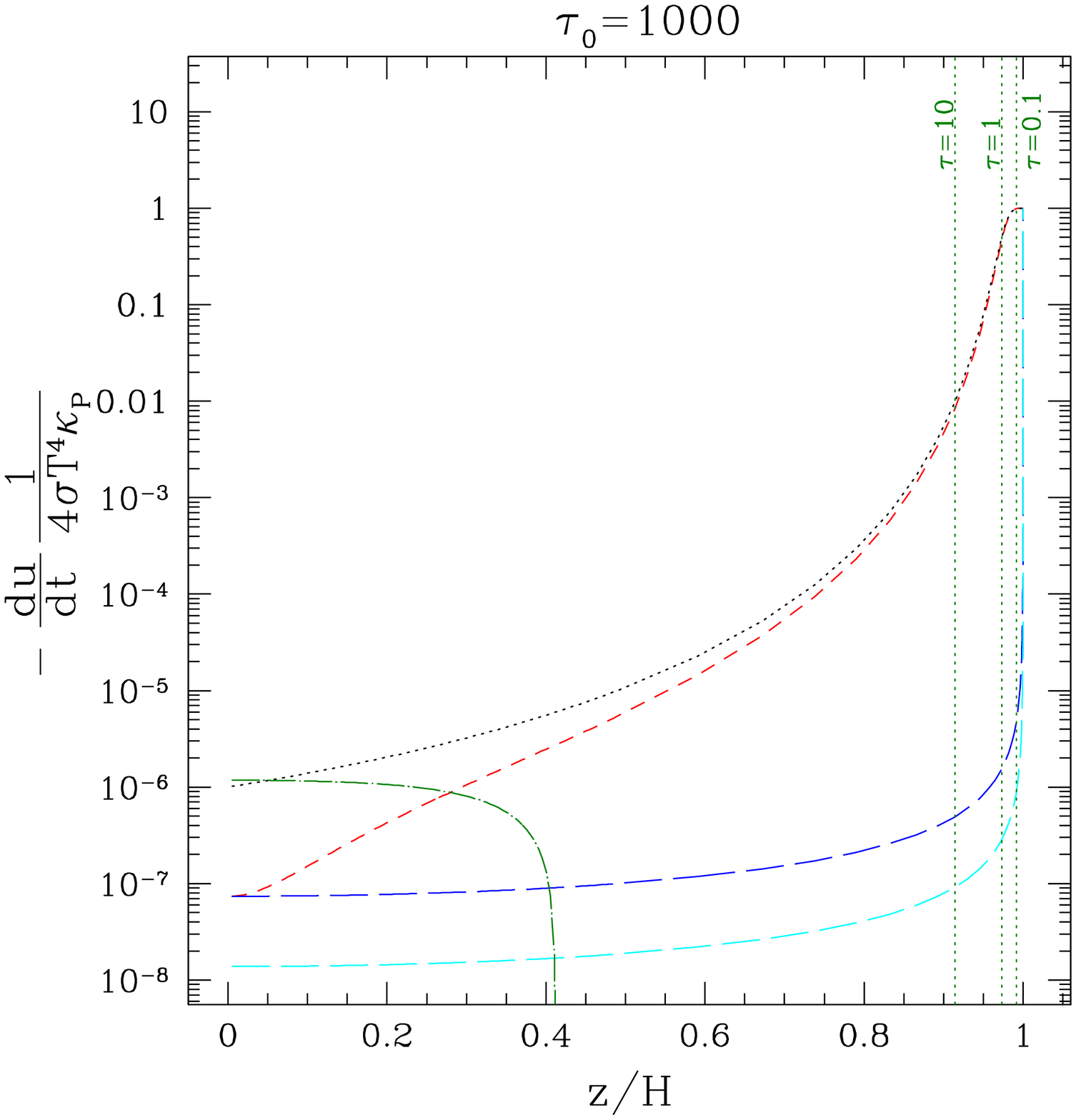}
	\includegraphics[width=0.84\columnwidth]{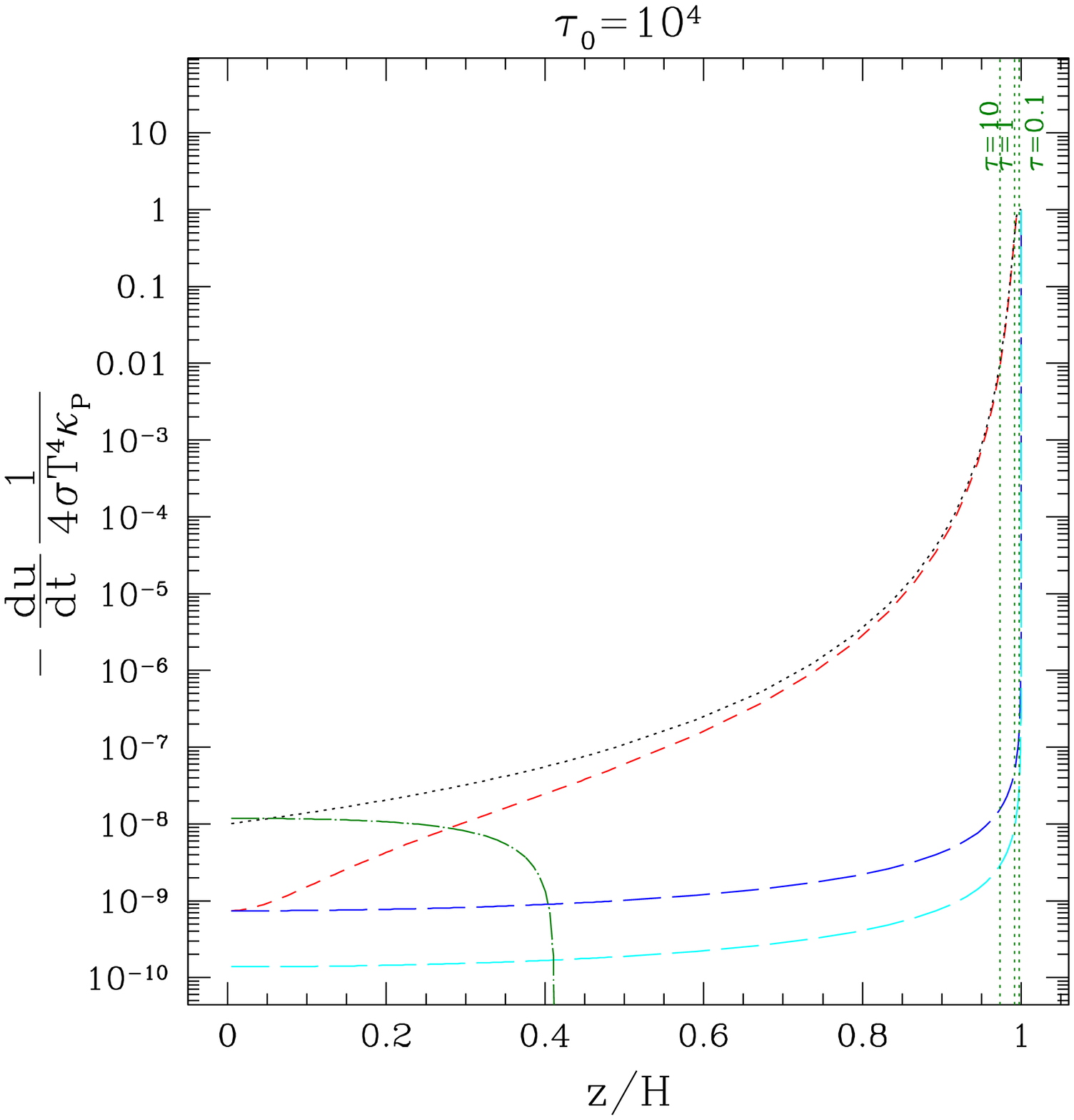}
	\caption{ The normalized radiative cooling rate $-du/dt$ versus normalized height $z$ in a disk of constant opacity and with a midplane optical depth $\tau_0$ of (a) 100, (b) 1000, and (c) $10^4$.  Line types are as in Figure \ref{fig:dudt}, with higher curves in Figure \ref{fig:dudt} corresponding to lower curves in this figure. Vertical green dotted lines show the locations where the actual optical depth $\tau_0$ is 0.01, 1, and 10.
} \label{fig:dudtdisk}
\end{figure}

Figure \ref{fig:dudtdisk} compares $du/dt$ at $R=5R_{\rm in}$ in a constant opacity disk as determined by equation (\ref{dudt}) in three different ways.  We again plot the cooling rate $-du/dt$ normalized to the optically thin limiting expression $4\sigma T^4 \kappa_{\rm P}$, and we note that all curves therefore approach 1 at the surface.
The dotted curve is the result for which the actual column density is used.  The red short dashed curve uses equation (\ref{barSigmai}) to estimate the column density, while the blue long dashed curve uses the estimate of SWBG. We choose $T_0=0$ and $\kappa_{\rm P}=\kappa_{\rm R}=\tau_0/\Sigma_0=521\tau_0\rho_{\rm in}^{-1}R_{\rm in}^{-1}$, where $\tau_0$ is the desired midplane optical depth and $\Sigma_0$ is the actual column density at the midplane.  The overestimate of column density in the SWBG method leads directly to an underestimate of the cooling rate.  We see that our method employing the pressure scale height is better able to reproduce the actual $du/dt$, with the agreement in the vicinity of the surface and photosphere improving as the midplane optical depth increases.

Also shown in Figure \ref{fig:dudtdisk}, as a green dot-dashed curve, is the diffusion limit result in the region $z/H\lap 0.41$, where the cooling rate $-du/dt$ is positive for the assumed temperature profile $T\propto P/\rho \propto \rho$.  We see here an example of the type of behavior dicussed at the end of \S3 of WC: the outer regions of the disk actually experience net {\it heating} due to their receiving more energy from the inner portion of the disk than they radiate outwards.  Fortunately, for those problems of interest in which the radiative diffusion timescale is much larger than the simulation time, accurately modelling radiative diffusion for arbitrary opacity laws and in the deep interior will not be important in comparison to modelling energy loss from the outer, optically thin fluid.  

\begin{figure}
	\centerline{\includegraphics[width=\columnwidth]{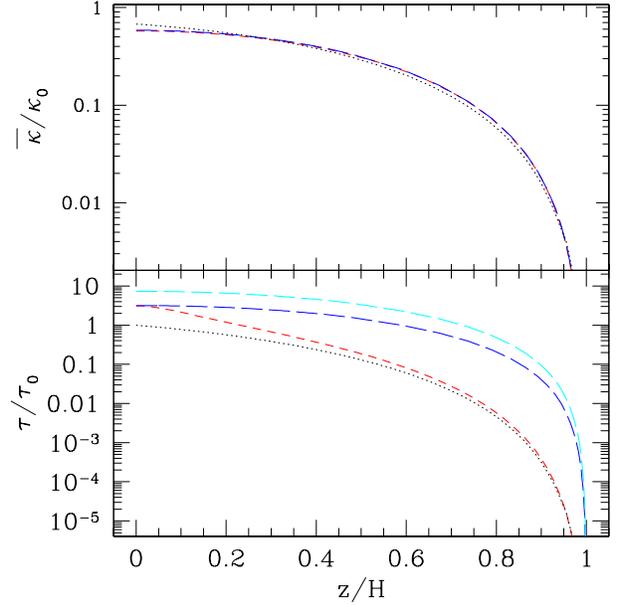}}
	\caption{Mean and pseudo-mean opacities normalized to the midplane opacity $\kappa_0$ (top panel) and optical depth normalized to the midplane optical depth $\tau_0$ (bottom panel), both as a function of $z$ for the same disk considered in Figure \ref{SigmaDisk} but with an ice grain opacity $\kappa \propto T^2$ and temperature $T\propto P/\rho \propto \rho$.  Line types are as in Figure \ref{SigmaDisk}.  Both the SWBG method and our method do a good job of estimating the actual mean $\bar \kappa$.  Given its accurate estimate of column density, our method yields an accurate estimate of optical depth throughout most of the disk. 
} \label{fig:kappatauice}
\end{figure}

As our final test scenario, we consider a disk with the same structure but now with an ice grain opacity of the form $\kappa \propto T^2$ and with the temperature $T \propto P/\rho \propto \rho$.   Thus, with the density distribution being described by equation (\ref{RhoDiskEquation}), the local opacity is given by $\kappa_{\rm R}=\kappa_0\left[2 \cos (z/H_0)-1\right]^2$, where $\kappa_0=\kappa_0(R)$ is the opacity at the midplane.  The optical depth $\tau$ in the $z$ direction can be determined analytically as
\begin{eqnarray}
\tau(R,z)&=&\kappa_0 \rho_{\rm in} \left(\frac{R_{\rm in}}{R}\right)^3 H_0  \left(\frac{9 \sqrt{3}}{2}-\frac{7 \pi }{3}+7 \frac{z}{H_0}\right. \nonumber\\
&&\left. -12 \sin\frac{z}{H_0}+3 \sin\frac{2 z}{H_0} -\frac{2}{3} \sin\frac{3 z}{H_0}\right).
\label{TauDiskEquation}
\end{eqnarray}
The actual mean opacity to the surface can then be obtained directly from equations (\ref{SigmaDiskEquation}) and (\ref{TauDiskEquation}) as $\bar\kappa=\tau/\Sigma$.  The pseudo-mean opacity for the ice grain case is just $\bar \kappa_{\rm R}=0.585 \kappa_{\rm R}$ for the original method of SWBG and $\bar \kappa_{\rm R}=0.575 \kappa_{\rm R}$ for our method.  The top panel of Figure \ref{fig:kappatauice} compares these mean opacities from $z$ to the surface at height $H$ as calculated in these three different ways.  Both the SWBG (long dashed curve) and our method (short dashed curve) agree well with the exact result (dotted curve).  The bottom panel of Figure \ref{fig:kappatauice} compares the corresponding optical depths from the same three calculation types, with the SWBG method yielding two curves, corresponding to the estimates done with and without the contribution from the central star to the gravitational potential.  The connection between the top and bottom panels is provided by the column density profiles shown in Figure \ref{SigmaDisk}.  The overestimate of column densities by the SWBG method leads directly to the overestimate of optical depth throughout the disk.  Although our method overestimates the central optical depth by about a factor of three, it does an excellent job of estimating the optical depth throughout most of the disk and especially near the surface.

\begin{figure}
	\includegraphics[width=0.84\columnwidth]{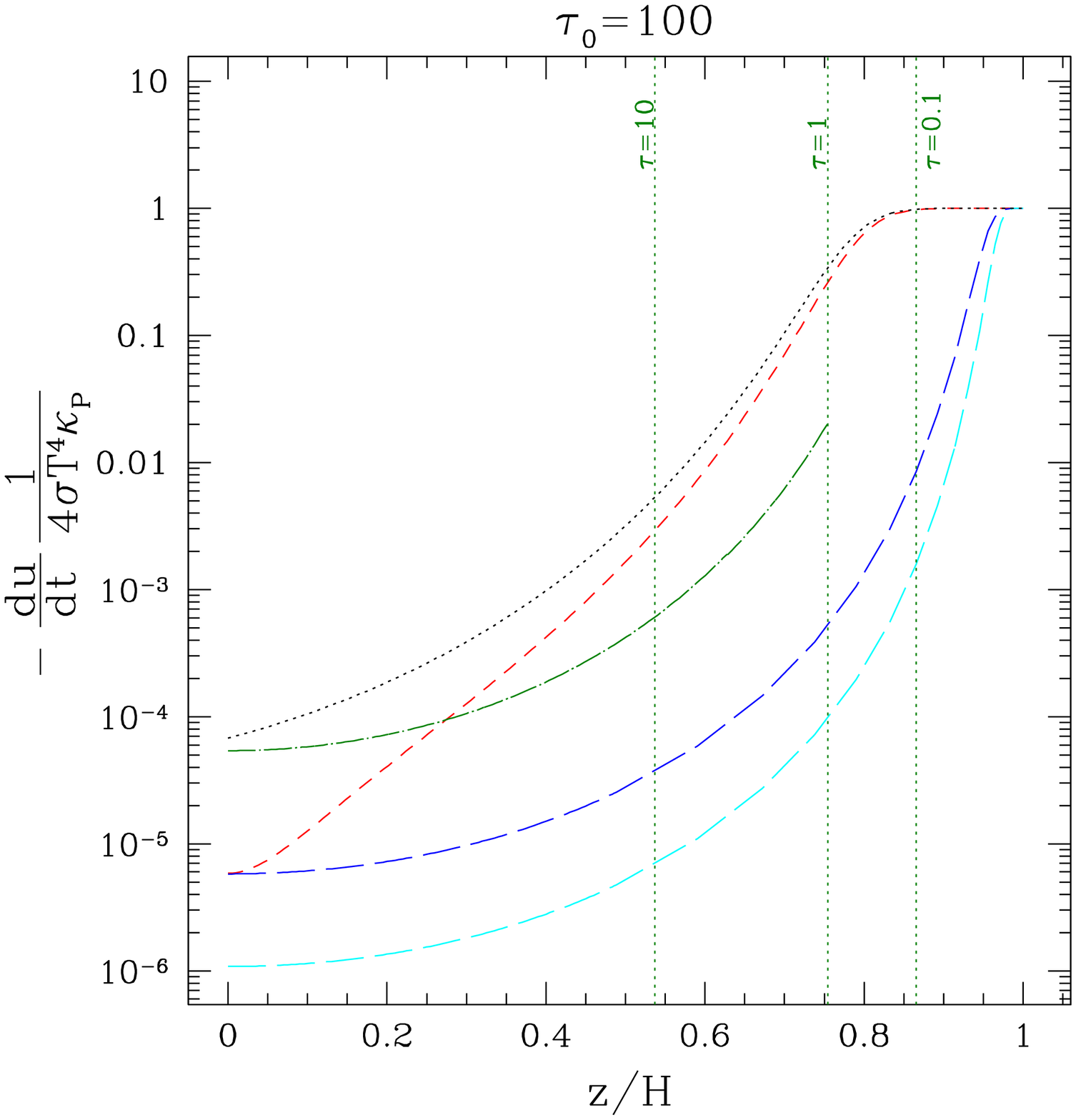}
	\includegraphics[width=0.84\columnwidth]{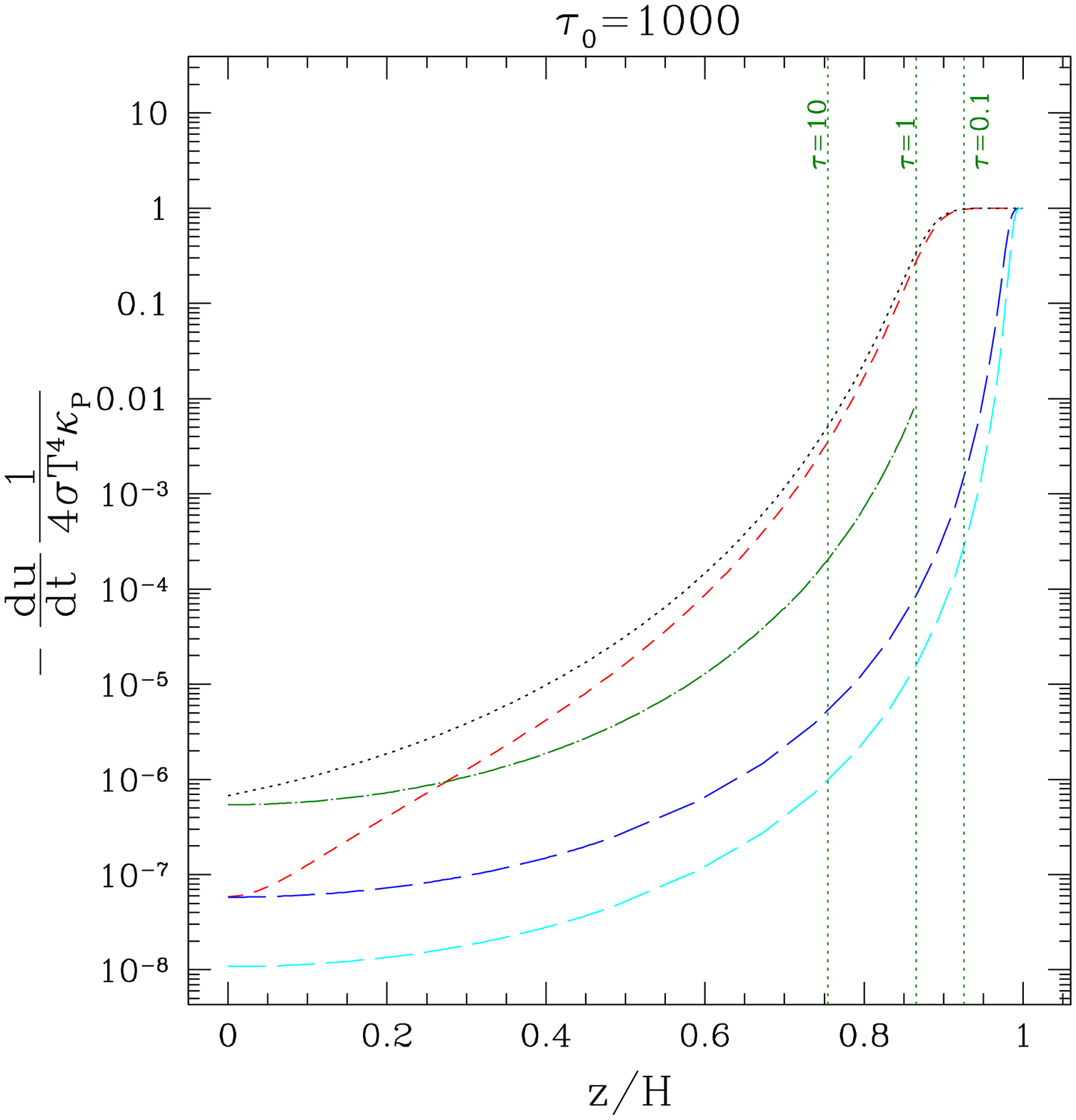}
	\includegraphics[width=0.84\columnwidth]{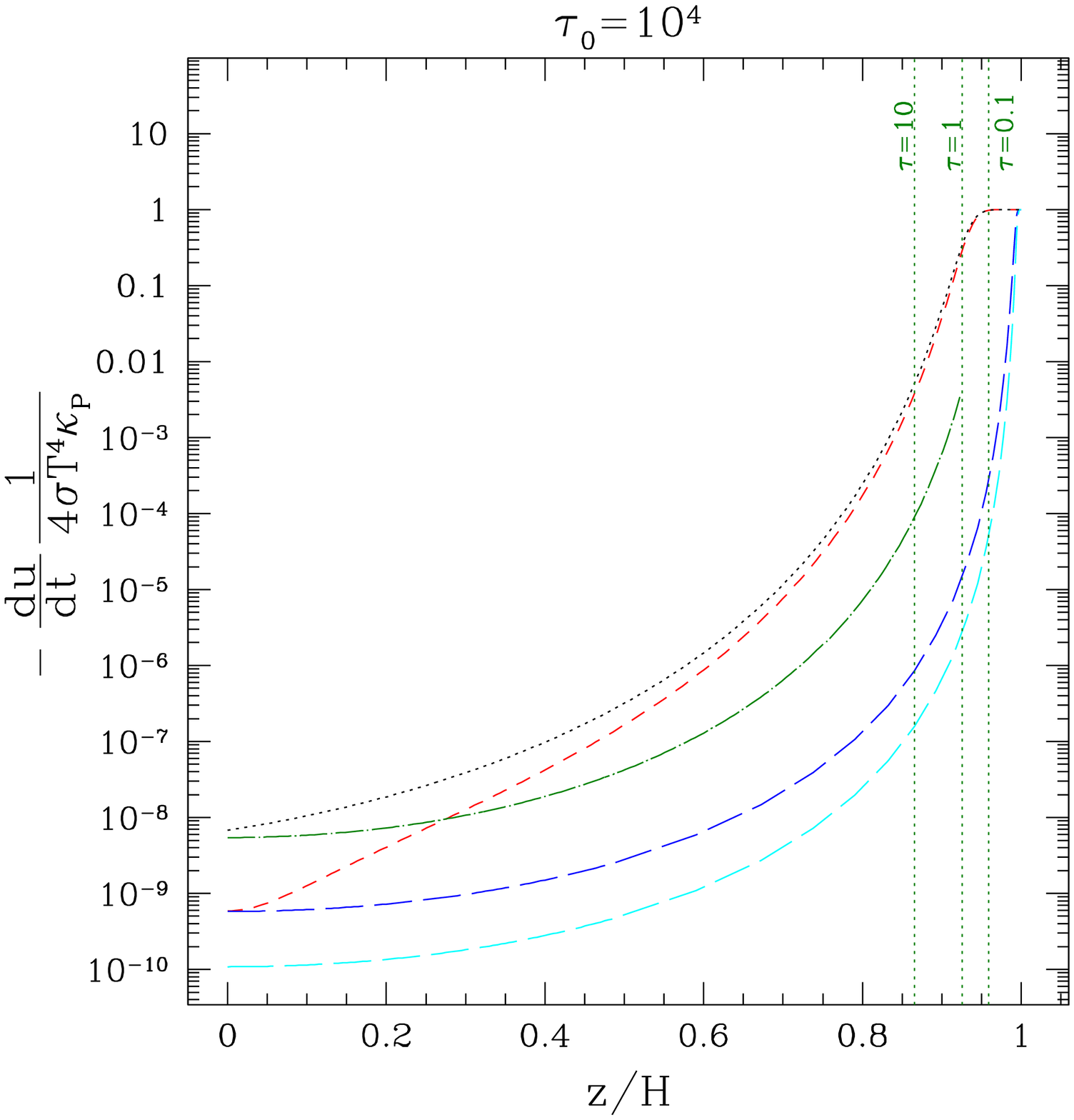}
	\caption{ The normalized radiative cooling rate versus normalized height at $R=5R_{\rm in}$ within disks as in Figure \ref{fig:kappatauice} with ice grain opacity.  The cases with midplane optical depth $\tau_0$ of (a) 100, (b) 1000, and (c) $10^4$ are shown.  Line types are as in previous figures.
%
%
} \label{fig:dudtdiskice}
\end{figure}

Figure \ref{fig:dudtdiskice} compares calculated $du/dt$ profiles at $R=5R_{\rm in}$ in the disks represented by Figure \ref{fig:kappatauice}.  This may be compared to Figure 7 in WC, although there the $\kappa_{\rm P}^{-1}$ term in the denominator of equation (\ref{dudt}) has been neglected.
As before, we plot the cooling rate $-du/dt$ normalized to the optically thin limiting expression $4\sigma T^4 \kappa_{\rm P}$.
The dotted curve is the result for which the actual column density is used.  The red short dashed curve uses equation (\ref{barSigmai}) to estimate the column density, while the blue long dashed curve uses the estimate of SWBG. We choose $T_0=0$ and $\kappa_{\rm P}=\kappa_{\rm R}$.  The midplane opacity $\kappa_0=770\tau_0\rho_{\rm in}^{-1}R_{\rm in}^{-1}$, where $\tau_0$ is the desired midplane optical depth.  We also include in Figure \ref{fig:dudtdiskice}, as a green dot-dashed curve, the diffusion limit result in the region of $\tau>1$.  As in the the case of the constant opacity disk, this diffusion limit result accounts for the radiative flux in both the $z$ and $r$ directions.  We note that the normalized diffusion limit cooling rate increases steadily outward from the midplane, unlike in the constant opacity case considered earlier in Figure \ref{fig:dudtdisk}.  Otherwise, Figure \ref{fig:dudtdiskice} reveals the same trends as in the constant opacity case, and our method continues to do an excellent job of estimating the cooling rate in regions of small and moderate optical depth.
Such behavior is to be expected, as the primary improvement from our method is a more accurate column density estimate, which is something independent of the particular opacity law being applied.

\section{Discussion}\label{sec:discussion}

Based on the tests presented here, the ``pressure scale height method'' retains the computational efficiency and accuracy of the original method proposed in SWBG for spherical geometries, while also providing a much better estimate of radiative cooling for disks and other fluid geometries.  In the end, the gravitational potential used in SWBG is a problematic tracer of the depth of a particle within a fluid configuration, because it is affected too much by long-range interactions that in no way contribute to the column density seen by a particle.  Furthermore, use of the gravitational potential as a tracer of depth makes some implicit assumptions about the closeness of a fluid configuration to equilibrium, since it cannot account for optically thick but otherwise low-mass regions in a configuration.  Pressure gradients, on the other hand, provide a direct estimate of depth within a fluid configuration and are the underlying effect responsible for stratification of density in the first place.  While the computation of gradients does require summation over neighbors, this is carried out within an SPH code regardless while calculating hydrodynamic forces and introduces no more computational time whatsoever.  The modified method should address many of the concerns raised in SWBG, WC, and other works about applying the method to situations other than star formation, particularly those involving stellar interactions.

Someone wishing to try our method in an existing code that implements the SWBG approach would simply need to use our equation (\ref{barSigmai}) with a constant $\zeta^\prime\approx 1.06$,
as opposed to SWBG's equation (18), to calculate the pseudo-mean column density $\bar \Sigma_i$ for use in determining $du_i/dt|_{\rm RAD}$.  In principle, one also needs to calculate pseudo-mean opacities according to equation (\ref{barkappaR}).  In practice, however, these pseudo-mean opacities never differ from those of SWBG by more than a few percent (see Figures \ref{pseudo} and \ref{fig:kappatauice}).  Therefore, existing codes that implement the SWBG method could continue to use the same pseudo-mean opacities, at least initially for the purpose of trying our method.

Given that other radiative transfer and cooling approaches are already available for grid-based Eulerian codes, we expect our method to be most immediately useful in SPH or other particle-based simulations.  However, the underlying idea in our method, or in the SWBG method for that matter,
is not specific to SPH.  Indeed, none of the tests of this paper require an SPH realization for the configurations, and the technique presented here could certainly be applied within grid-based
codes as well: in this case, the index $i$ would refer not to a particle but to a grid cell instead.

While our modified method, like the original method of SWBG, loses accuracy at large optical depths, it is completely compatible with the techniques described in \citet{2009MNRAS.394..882F} which add a flux-limited diffusion term to the cooling equations representing essentially a radiative thermal conductivity between neighboring regions.  As shown there, such techniques generally add a small relative fraction to the total simulation time, and may be easily incorporated within SPH schemes via particle-neighbor interaction terms.  Whether or not such terms are likely to have much of an effect on the results of a simulation comes down to a timescale issue: if the thermalization timescale within a configuration is short relative to the local hydrodynamic timescale, the interaction terms should provide an important contribution to energy transfer; otherwise, hydrodynamical effects will serve as the primary mechanism instead.

Should it become necessary, one could develop a more robust crossover method that smoothly connects the actual diffusion result  $du/dt|_{\rm thick}$ (implemented with a flux limiter) and the optically thin result $du/dt|_{\rm thin}$.  A local estimate of the optical depth $\tau$, calculated via equation (\ref{barkappai}), could help provide an appropriate weighting factor.  Along these lines, we have begun to experiment with $du/dt|_{\rm RAD}=\exp(-\tau) du/dt|_{\rm thin} + [1 -\exp(-\tau)] du/dt|_{\rm thick}$.  Comparison with three-dimensional hydrodynamic simulations with full radiative transport would further help to refine and calibrate such implementations.


\section*{Acknowledgments}

We thank D.\ Stamatellos for helpful email exchanges and S.\ Rahman for useful discussions.  This
work was supported by the National Science Foundation (NSF) grant numbers AST-1313091 and PHY-1212426, and it used the Extreme Science and
Engineering Discovery Environment (XSEDE), which is supported by NSF grant
number OCI-1053575.

\bsp

\label{lastpage}

\end{document}